\documentclass[a4paper,11pt]{article}
\usepackage{amsmath}
\usepackage{amsthm}
\usepackage{amsfonts}
\usepackage{mathrsfs}
\usepackage{amssymb}
\usepackage{mathpazo}
\usepackage{mathabx}
\usepackage{accents}
\usepackage[dvipsnames]{xcolor}
\usepackage{bm}
\usepackage[no-weekday]{eukdate}
\usepackage[bb=boondox]{mathalfa}
\usepackage{paralist}
\usepackage{natbib}
\usepackage{url}
\usepackage[textwidth=6.275in,textheight = 9.5in]{geometry}
\usepackage{placeins}
\usepackage[hidelinks]{hyperref}
\hypersetup{
	pdftitle = {{Forecast reconciliation with non-linear constraints}}, 
	pdfauthor = {{Daniele Girolimetto, Anastasios Panagiotelis, Tommaso Di Fonzo, and Han Li}}
}
\usepackage{doi}

\usepackage{tikz}
\usetikzlibrary{calc, fit, backgrounds}
\newcommand{\relation}[3]
{
	\draw (#3.south) -- +(0,-#1) -| ($ (#2.north) $)
}

\usepackage{algorithm}     
\usepackage{algpseudocode} 
\usepackage{multicol}

\def\equationautorefname~#1\null{Eq.~(#1)\null}

\usepackage{multirow, float}
\usepackage{makecell}
\usepackage{calc}
\usepackage{graphicx}
\usepackage{enumitem}
\usepackage{microtype}
\usepackage{caption}
\DeclareCaptionStyle{italic}[justification=centering]
 {labelfont={bf},textfont={it},labelsep=colon}
\usepackage[compact]{titlesec}
\titlespacing{\section}{0pt}{2ex}{1ex}
\titlespacing{\subsection}{0pt}{1ex}{0ex}
\titlespacing{\subsubsection}{0pt}{0.5ex}{0ex}
\usepackage[bottom,hang,flushmargin]{footmisc}
\usepackage{adjustbox}

\allowdisplaybreaks
\newcommand{\blind}{1}

\usepackage{caption}
\usepackage{subcaption}

\usepackage{longtable}
\usepackage{booktabs}
\usepackage{array}
\newcolumntype{M}[1]{>{\centering\arraybackslash}m{#1}}
\newcolumntype{L}[1]{>{\raggedright\arraybackslash}m{#1}}
\newcolumntype{R}[1]{>{\raggedleft\arraybackslash}m{#1}}

\newcommand{\deltavet}{\bm{\delta}}

\newcommand{\lambdavet}{\bm{\lambda}}

\newcommand{\dvet}{\bm{d}}
\newcommand{\evet}{\bm{e}}
\newcommand{\gvet}{\bm{g}}

\newcommand{\jvet}{\bm{j}}

\newcommand{\vvet}{\bm{v}}

\newcommand{\yvet}{\bm{y}}
\newcommand{\zvet}{\bm{z}}

\newcommand{\Bvet}{\bm{B}}

\newcommand{\Ivet}{\bm{I}}
\newcommand{\Jvet}{\bm{J}}

\newcommand{\Wvet}{\bm{W}}

\newcommand{\Yvet}{\bm{Y}}

\newcommand{\Zerovet}{\bm{0}}

\newcommand{\Sigmavet}{\bm{\Sigma}}

\newcommand{\muvet}{\bm{\mu}}
\newcommand{\kappavet}{\bm{\kappa}}

\DeclareRobustCommand{\widebreve}[1]{\accentset{\smile}{#1}}

\newcommand{\yhat}{\widehat{\yvet}}
\newcommand{\ystar}{\yvet^*}
\newcommand{\ytilde}{\widetilde{\yvet}}
\newcommand{\ybreve}{\widebreve{\yvet}}
\newcommand{\cM}{\mathcal{M}}

\definecolor{mybluehl}{HTML}{cbd3ff}

\makeatletter
\newcommand{\githuburl}{\begingroup%
\if0\blind
{
\texttt{<url not available for blind revision>}
}\fi
\if1\blind
{
\url{https://github.com/danigiro/nlcr}.
}\fi
\endgroup}
\makeatother

\makeatletter
\newcommand{\pkgname}{\begingroup%
\if0\blind
{
\texttt{<name not available for blind revision>}
}\fi
\if1\blind
{
\texttt{nlcReco}
}\fi
\endgroup}
\makeatother

\makeatletter
\def\@endtheorem{\endtrivlist}
\makeatother

\theoremstyle{definition}

\newtheorem{theorem}{Theorem}[section]
\newtheorem{lemma}{Lemma}[section]

\usepackage[affil-it]{authblk}
\makeatletter
\renewenvironment{abstract}{%
    \if@twocolumn
      \section*{\abstractname}%
    \else %
      \begin{center}%
        {\bfseries \large\abstractname\vspace{\z@}}%
      \end{center}%
      \quotation
    \fi}
    {\if@twocolumn\else\endquotation\fi}
\makeatother

\title{Forecast reconciliation with non-linear constraints}

\author[1]{Daniele Girolimetto}
\author[2]{Anastasios Panagiotelis}
\author[1]{Tommaso Di Fonzo}
\author[4]{Han Li}

\affil[1]{\small Department of Statistical Sciences, University of Padova, Padova 35121, Italy}
\affil[2]{\small Department of Econometrics and Business Statistics, Monash University, VIC 3800, Australia}
\affil[4]{\small Department of Economics, The University of Melbourne, VIC 3010, Australia}

\date{\today}

\makeatletter
\ifnum\blind=0
  \renewcommand\maketitle{%
    \begin{center}
      {\LARGE\bfseries \@title\par}
      \vskip1em
      {\large \@date\par}
    \end{center}
    \vskip2em
  }
\fi
\makeatother

\begin{document}

\def\spacingset#1{\renewcommand{\baselinestretch}{#1}\small\normalsize}
\spacingset{1.1}

\thispagestyle{empty}\clearpage\maketitle

\ifnum\blind=1
{
\begingroup
\let\thefootnote\relax\footnotetext{\raggedright Emails: \href{mailto:daniele.girolimetto@unipd.it}{\texttt{daniele.girolimetto@unipd.it}}, \href{mailto:anastasios.panagiotelis@monash.edu}{\texttt{anastasios.panagiotelis@monash.edu}}, \href{mailto:difonzo@stat.unipd.it}{\texttt{difonzo@stat.unipd.it}}, and \href{mailto:han.li@unimelb.edu.au}{\texttt{han.li@unimelb.edu.au}}}
\endgroup
}\fi

\begin{abstract}
\noindent Methods for forecasting time series adhering to linear constraints have seen notable development in recent years, especially with the advent of forecast reconciliation. This paper extends forecast reconciliation to the open question of non-linearly constrained time series. Non-linear constraints can emerge with variables that are formed as ratios such as mortality rates and unemployment rates. On the methodological side, Non-linearly Constrained Reconciliation (NLCR) is proposed. This algorithm adjusts forecasts that fail to meet non-linear constraints, in a way that ensures the new forecasts meet the constraints. The NLCR method is a projection onto a non-linear surface, formulated as a constrained optimisation problem. On the theoretical side, optimisation methods are again used, this time to derive sufficient conditions for when the NLCR methodology is guaranteed to improve forecast accuracy. Finally on the empirical side, NLCR is applied to two datasets from demography and economics and shown to significantly improve forecast accuracy relative to relevant benchmarks.
\end{abstract}

\begin{itemize}[nosep, align=left, leftmargin = !]
	\item[\textbf{Keywords}] \textit{Forecast reconciliation,~~Non-linear constraints,~~Hierarchical forecasting,~~Optimisation,~~Mortality,~~Unemployment}
\end{itemize}

\vfill

\newpage
\spacingset{1.3}


\section{Introduction}\label{sec:intro}

Many problems in operations research involve forecasting multiple variables that meet certain constraints. These constraints are often linear, arising from aggregation. For example, decision making may depend on forecasts of demand across a supply chain where store level demand aggregates to the level of demand in a regional warehouse, which in turn aggregates to overall demand in the organisation. While data will cohere to these aggregation structures, forecasts made by different agents will typically be incoherent, that is, they fail to meet the constraints seen in the data. Forecast reconciliation has emerged as a post-forecasting process that adjusts incoherent ``base'' forecasts to ensure they are coherent, thus facilitating aligned decision making, while also improving the accuracy of forecasts. A comprehensive review of forecast reconciliation can be found in \cite{athanasopoulos2024}. To date, there has been little to no consideration on the case where time series are connected by \emph{non-linear} constraints; a gap that we address in this paper.

For the case of linear constraints, ideas of reconciliation date back to the work of \cite{stone1942} and \cite{byron1978, byron1979}, although without an emphasis on forecasting. Within the field of forecasting, for some time the extant approach was to only forecast variables at a single level, e.g. the top down and bottom up approaches \citep{gross1990}. These approaches ignore the operational reality of many settings where forecasts are available at all levels each encapsulating important information. In a series of papers \citep{athanasopoulos2009,hyndman2011,wickramasuriya2019} forecast reconciliation methods were developed that exploit base forecasts of all variables under consideration rather than only those at a single level. These were initially motivated through a regression modelling approach. 

The later work of \cite{panagiotelis2021} reinterpreted forecast reconciliation through the lens of optimisation and geometry. It is this approach that most directly inspires our new work on non-linear constraints, to be developed below in detail. In the setting of linear constraints \cite{panagiotelis2021} discuss how in forecast reconciliation, base forecasts (denoted $\widehat{\bm y}$) belong to a domain (e.g. $\mathbb{R}^n$), while the observations (denoted $\yvet$) belong to a linear subspace of this domain  $\mathcal{S}\subset\mathbb{R}^n$, referred to as the `coherent subspace'. As such, forecast reconciliation involves \emph{mapping} base forecasts to the coherent subspace to obtain reconciled forecasts $\widetilde{y}=\psi(\widehat{\bm y})$ where $\psi:\mathbb{R}^n\rightarrow\mathcal{S}$. One possible mapping is a projection. A projection solves an optimisation problem in that it minimises the distance between $\widehat{\bm y}$ and $\widetilde{\bm y}$ (denoted $d(\widehat{\bm y},\widetilde{\bm y})$) subject to $\widetilde{\bm y}\in\mathcal{S}$. By `distance' we refer to Euclidean distance, although it is common in reconciliation to rescale the variables in a way that exploits the variance and covariances of base forecasting errors. Using these notions a key result of \cite{panagiotelis2021} is that reconciliation is guaranteed to reduce forecast error in the sense that $d({\bm y},\widetilde{\bm y})\leq d({\bm y},\widehat{\bm y})$.

A shortcoming of the current literature is that only linear constraints are considered. However non-linear constraints arise in a number of interesting applications. For example, many important variables that need to be forecast are \emph{ratios}. Two important examples we consider in empirical work are the case of mortality rates (defined as number of deaths divided by population exposure) and unemployment rates (number of unemployed divided by labour force participation). For ratios, it is possible to log-linearise and then apply linear reconciliation methods. However, for problems with ratios, some variables may themselves be subject to other linear constraints. For example, the number of unemployed or number of deaths in a country can be disaggregated by geography, age or gender. In this setting, with a full system of linear and non-linear constraints, it is no longer possible to linearise the ratio constraint without inducing non-linearity in the aggregation constraints. Furthermore there are other examples where forecasts need to be made of variables that are related to one another via non-linear constraints, for example in wind power applications, the power curve is a non-linear function of the windspeed \citep{messner2014, xu2016}, providing another potential use case of the methods we develop in this paper. Other motivating examples can also be found in the case of temporal aggregation for models that employ non-linear transformations \citep[for example see][]{proietti2006, proietti_moauro2006}

Having motivated the need for non-linear methods of reconciliation, we now turn our attention to the novel contributions of this paper. The first is to extend the notions of coherence and reconciliation via projections to the non-linear setting. Unlike the linear setting, for general non-linear constraints, there will not be a closed form solution for the mapping $\psi$. However we show that projection can be achieved as a solution to an optimisation problem solved using the Lagrangian method. We then turn our attention to the theoretical properties of this projection. An important insight is that unlike for the linear case, there is no guarantee that $d({\bm y},\widetilde{\bm y})\leq d({\bm y},\widehat{\bm y})$ when constraints are non-linear. We therefore turn our attention to find conditions where this does hold. Again we employ an optimisation approach, finding a critical point that is equidistant from the base and reconciled forecast. This critical point, together with the reconciled forecasts form a ball, such that for any realisation within this ball, reconciliation is guaranteed to improve forecast accuracy. We rigorously derive an expression for the radius of the ball, which depends on the gradients of the constraint functions and Lagrange multipliers of optimisation problems described in detail below. From these results we are able to make practical conclusions that guide forecasters on knowing when our proposed non-linearly constrained forecast reconciliation approach is more likely to succeed. Factors that are influential include the curvature of the constraints, the distance of the base forecast from the coherent subspace, and whether any constraints are convex. Further illustration on the theoretical results is demonstrated in an extensive simulation study.

Finally, we apply our newly developed non-linearly constrained forecast reconciliation methods to two empirical case studies. The first is US mortality rates, with deaths and exposure broken down by age and geography. The second is Australian unemployment rates broken down by geography. As benchmarks we consider (i) forecasting the ratio alone and (ii) taking the ratio of forecasts of the numerator and denominator (analogous to a bottom up approach). For both datasets we find that our proposed non-linearly constrained forecast reconciliation method improves forecast accuracy relative to these benchmarks. When the variables are suitably rescaled to account for base forecast error covariance, these improvements over benchmarks are statistically significant in both empirical applications. The code and data for reproducing the results are available at \githuburl

The remainder of the paper is structured as follows. Section~\ref{sec:theory} introduces the method as well derives the theoretical results on when reconciliation is guaranteed to improve forecast accuracy. Section~\ref{sec:alg} provides implementation details including those relevant to the empirical work in this paper. Section~\ref{sec:sim} is an extensive simulation study that sheds further light on the theoretical results. Section~\ref{sec:app} includes both empirical studies and Section~\ref{sec:conclusion} concludes.

\section{Reconciliation for non-linear constraints}\label{sec:theory}

Let $\Yvet$ be a random $n$-vector with realisations $\yvet$ that meet $C<n$ constraints $g_1(\yvet)=0,g_2(\yvet)=0,\dots,g_C(\yvet)=0$. We can write this compactly as $\gvet(\yvet)=\Zerovet$ where $\gvet:\mathbb{R}^n\rightarrow\mathbb{R}^C$. The constraint function may consist of both linear and non-linear constraints. The case where all constraints are linear is well studied, therefore we focus on the case where there is at least one non-linear constraint. We assume all constraint functions to be continuous functions.

The level set of points $\yvet:\gvet(\yvet)=\Zerovet$ is a manifold $\cM$ that we refer to as the $\textit{coherent manifold}$. All realisations belong to the coherent manifold, $\yvet\in\cM$. 

\subsection{Methodology}\label{sec:opt2}

Let $\yhat\in\mathbb{R}^n$ be a \textit{base} forecast (prediction) that is incoherent, i.e., $\yhat\notin\cM$. We propose to find a reconciled forecast $\ytilde\in\cM$ by solving the optimisation problem
\begin{align}
	\ytilde =& \underset{\zvet}{argmin}(\zvet - \yhat)'\bm{W}^{-1}(\zvet - \yhat)\label{eq:obj}\\
	\textrm{s.t.}&\, \gvet(\zvet)=\Zerovet\label{eq:objc}
\end{align}
where $\bm{W}$ is a positive definite matrix. For example, motivated by the MinT method \citep{wickramasuriya2019} $\bm{W}$ may be an estimate of the covariance matrix of base forecast errors.

The optimisation problem in Equations~(\ref{eq:obj}) and~(\ref{eq:objc}) generalises existing methods for linear constraints by `projecting' $\yhat$ to the nearest point on the coherent manifold. To solve, we formulate the problem as an unconstrained optimisation with Lagrangian
\begin{equation}
	\mathcal{L} = (\zvet - \yhat)'\bm{W}^{-1}(\zvet - \yhat) - 2\lambdavet' \gvet(\zvet)
\end{equation}
where $\lambdavet$ is a $C$-vector of Lagrange multipliers. The gradient of the Lagrangian
is
\begin{align}
	\frac{\partial\mathcal{L}}{\partial\zvet}&=2\bm{W}^{-1}(\zvet-\yhat) - 2\Jvet\lambdavet\label{eq:focz}\\
	\frac{\partial\mathcal{L}}{\partial\lambdavet}&=\gvet(\zvet)
\end{align}
where $\Jvet$ is the Jacobian
\[
\Jvet=\begin{pmatrix}
\displaystyle	\frac{\partial g_1(\zvet)}{\partial z_1}&\dots&
\displaystyle	\frac{\partial g_C(\zvet)}{\partial z_1}\\
	\vdots&\ddots&\vdots\\
\displaystyle	\frac{\partial g_1(\zvet)}{\partial z_n}&\dots&
\displaystyle	\frac{\partial g_C(\zvet)}{\partial z_n}\\
\end{pmatrix} .
\]
The solution is found at
$\yvet=\ytilde$
meaning that \autoref{eq:focz} implies
\begin{equation}
	\ytilde=\yhat + \bm{W}\bm{\widetilde J}\lambdavet\label{eq:reco_sol} ,
\end{equation}
where $\widetilde{\bm J}$	is the Jacobian evaluated at
$\yvet=\ytilde$.

This can be interpreted as follows. The products of the rows of matrix $\bm{W}$ and the columns of $\widetilde{\Jvet}$ (the gradients of the constraint functions) provide a basis for a linear subspace along which $\yhat$ is `projected' onto $\ytilde$. The Lagrange multipliers measure `how far' to project along each basis direction, while the signs of the Lagrange multipliers indicate the directions of projection.

\subsection{Guarantees on reconciliation improving accuracy}\label{sec:guar}

We are interested in the case where reconciliation improves forecast accuracy compared to the base forecast, where accuracy is in a mean squared error sense. More formally, reconciliation improves forecast accuracy when $d(\yvet,\ytilde)-d(\yvet,\yhat)<0$. For simplicity of exposition, in the remainder of this Section we consider $\bm{W}=\bm{I}$, in which case $d()$ is Euclidean distance. If $\bm{W}\neq\bm{I}$ $d()$ is Mahalanobis distance \citep{mahalanobis1936}, i.e.,  the forecast accuracy is in terms of a rescaled version of Euclidean distance. Our main result is to find a critical point $\breve{\yvet}$. A ball with $\ytilde$ at the center $\breve{\yvet}$ on its exterior is then found such that $d(\yvet,\ytilde)-d(\yvet,\yhat)<0$ for all $\yvet$ inside the ball.

We first construct a separating hyperplane. This hyperplane separates all points in $\mathbb{R}^n$ into two sets, the first consisting of all points closer to $\ytilde$, the second consisting of all points closer to $\yhat$. This hyperplane $\mathscr{H}$ is orthogonal to the line $\mathscr{L}$ connecting $\yhat$ and $\ytilde$, with the latter parallel to the vector $\widetilde{\Jvet}\lambdavet$. The $\mathscr{H}$ and $\mathscr{L}$ intersect at the midpoint of $\mathcal{L}$, given by $\ytilde+\frac{1}{2}\widetilde{\Jvet}\lambdavet$. The hyperplane is thus given by

\[\yvet'\widetilde{\Jvet}\lambdavet-c=0\]

\noindent where
$c=\ytilde'\widetilde{\Jvet}\lambdavet + \frac{1}{2}\lambdavet'\widetilde{\Jvet}'\widetilde{\Jvet}\lambdavet$. We can find a region of $\yvet$ for which reconciliation guarantees forecast improvement, by finding the point nearest to $\ytilde$ where $\cM$ intersects $\mathscr{H}$. This is found by solving the optimisation problem

\begin{align}
	\ybreve =& \underset{\vvet}{argmin}(\ytilde-\vvet)'(\ytilde-\vvet)\label{eq:obj2}\\
	\textrm{s.t.}&\, \gvet(\vvet)=\Zerovet\\
	&\vvet'\widetilde{\Jvet}\lambdavet-c=0
\end{align}
which has a Lagrangian 
$$
\mathcal{L} = (\ytilde-\vvet)'(\ytilde-\vvet)+\kappavet' \gvet(\vvet)+\mu(\vvet'\widetilde{\Jvet}\lambdavet-c)
$$
with gradient
\begin{align}
	\frac{\partial\mathcal{L}}{\partial\vvet}=&2(\vvet-\ytilde)+ \Jvet\kappavet\label{eq:focz2}+\mu\widetilde{\Jvet}\lambdavet\\
	\frac{\partial\mathcal{L}}{\partial\kappavet}=&\gvet(\vvet)\\
	\frac{\partial\mathcal{L}}{\partial\mu}=& \vvet'\widetilde{\Jvet}\lambdavet-c
\end{align}
The solution is found at $\vvet=\ybreve$ meaning that \autoref{eq:focz2} implies
\begin{equation}
	\ybreve=\ytilde-
\frac{1}{2}\widebreve{\Jvet}\kappavet
-\frac{\mu}{2}\widetilde{\Jvet}\lambdavet\label{eq:reco_sol2} ,
\end{equation}
where $\widebreve{\bm J}$	is the Jacobian evaluated at $\bm {v}=\breve{\yvet}$ multiplied by a scaling factor of 1/2.
The distance between $\ytilde$ and $\widebreve{\yvet}$ is given by
\begin{equation}
	r=\sqrt{\kappavet'\widebreve{J}'\widebreve{J}\kappavet+\mu\kappavet'\widebreve{J}'\widetilde{J}\lambdavet+\frac{\mu^2}{4}\lambdavet\widetilde{J}'\widetilde{J}\lambdavet}\label{eq:radius}
\end{equation}

\begin{theorem}\label{thm:thm1}
	Consider the ball $\mathcal{B}_{\ytilde,r}$ where $r$ is the radius given in \autoref{eq:radius}. Then $d(\ytilde,\yvet)<d(\yhat,\yvet)\,\forall \yvet\in\cM$ is guaranteed whenever $\Yvet\in\mathcal{B}_{\ytilde,r}\cap\cM$
\end{theorem}
\begin{proof}
	Let $f(\yvet):\cM\rightarrow\mathbb{R}$ be the function $f(\yvet)=d(\yvet,\ytilde)-d(\yvet,\yhat)$. This measures gain in forecasting accuracy that accrues from reconciliation for a given value of $\yvet\in\cM$. Note that $f(\cdot)$ will be a continuous function as long as $g(\cdot)$ is continuous. Also, since $\yhat\notin\cM$ $f(\ytilde)<0$ with strict inequality. Now, suppose there exists some point $\yvet^{\dagger}$ in the set $\mathcal{B}_{\ytilde,r}\cap\cM$ such that $f(\yvet^{\dagger})$ is a small positive value $f(\yvet^{\dagger})=\epsilon>0$. By the continuity of $f(\cdot)$ this implies that there must be a point arbitrarily close to $\yvet^{\dagger}$ for which $f(\yvet)=0$. However, such a point would satisfy the constraints of the optimisation problem in \autoref{eq:obj2} yet be even closer to $\ytilde$ than $\ybreve$. Since $\ybreve$ is the minimiser of the optimisation problem, this leads to proof by contradiction.
\end{proof}

\subsubsection{Special case 1: One constraint}

In the one constraint case, the Jacobian is a vector which we denote by $\widetilde{j}$ and $\widebreve{j}$ when evaluated at $\widetilde{y}$ or $\widebreve{y}$ respectively.  \autoref{eq:radius} simplifies to

\[	
r=\sqrt{\kappa^2\langle\widebreve{j},\widebreve{j}\rangle+\mu\kappa\lambda\langle\widebreve{j},\widetilde{j}\rangle+\frac{\mu^2}{4}\lambda^2\langle\widetilde{j},\widetilde{j}\rangle}
\]

\begin{figure}[tb]
	\centering
\includegraphics[trim=2cm 2cm 0.8cm 3cm, clip, width=0.7\linewidth]{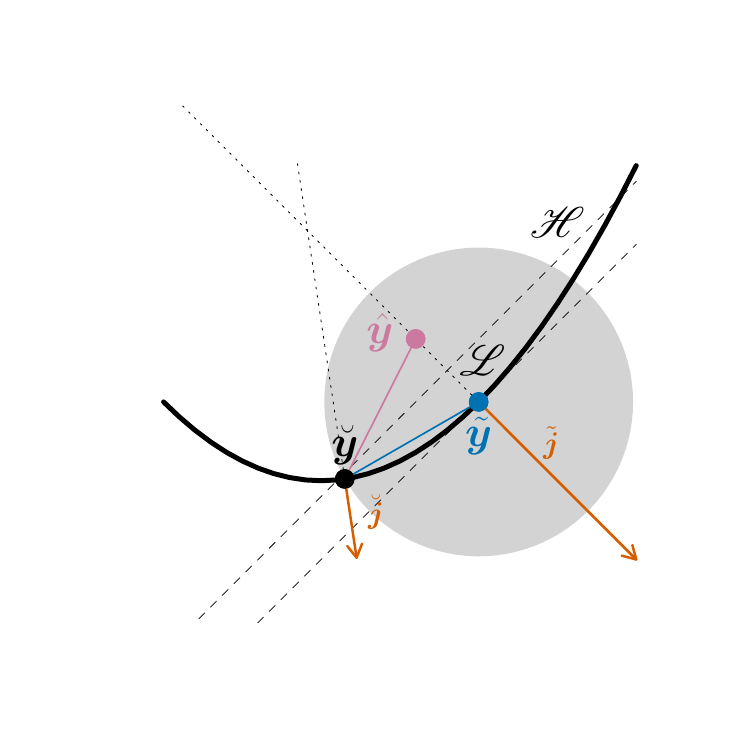}
	\caption{Schematic demonstrating optimisation problem described in \autoref{sec:opt2}. The point $\ybreve$ will lie at a point where $\cM$ intersects with a hyperplane that bisects $\yhat$ and $\ytilde$, shown as a dotted line. The gradient $\nabla g(\zvet)$ evaluated at $\ytilde$ and $\ybreve$ is denoted as $\widetilde{\jvet}$ and $\widebreve{\jvet}$, respectively. The lengths of the dotted lines will influence the values of the Lagrange multipliers $\mu$ and $\kappa$. The shaded grey region is the ball, such that when realisations are inside the ball, reconciliation is guaranteed to improve accuracy relative to base forecasts. }
	\label{fig:qual}
\end{figure}

\autoref{fig:qual} depicts a situation where $\langle\widebreve{j},\widetilde{j}\rangle>0$, $\lambda<0$, $\kappa<0$ and $\mu>0$. Recall that reconciliation is guaranteed to improve forecast accuracy when realisations are inside the ball with radius $r$. Therefore reconciliation is more effective when, other things being equal, the radius of the ball is larger. This occurs when

\begin{itemize}
	\item $\lambda$ and $\kappa$ are larger (in absolute value), i.e. when $\yhat$ is further away from $\cM$.
	\item When the inner products $\langle\widebreve{j},\widetilde{j}\,\rangle$, $\langle\widetilde{j},\widetilde{j}\,\rangle$, $\langle\widebreve{j},\widebreve{j}\,\rangle$ are larger. This in turn is related to the rate of change of the gradient over the constraint function. If the gradient is stable over a larger region of the constraint then the ball will have a larger radius. The Gaussian curvature is equal to the instantaneous rate at which a tangent vector rotates \citep{pressley2010}, a quantity approximated up to proportionality by $\langle\widebreve{j},\widetilde{j}\rangle$ when $\yhat$ and $\ytilde$ are close. As the constraint approaches no curvature (a linear constraint) the ball becomes larger. This is in line with results for linear constraints, where reconciliation always leads to improvements in forecast accuracy.  $\accentset{\smile}{j},$
\end{itemize}

\subsubsection{Special case 2: Convex constraints}\label{sec:convex}

We now consider the special case where each $g_c(\cdot)$ is a \textit{convex} function. Alternatively, we can deal with functions that are not globally convex by restricting our attention to a high-probability region of $\cM$ over which each $g_c(\cdot)$ is convex. 

We first consider the case of one constraint, $g(\zvet)=0$ in which case \autoref{eq:reco_sol} simplifies to

\[
\ytilde=\yhat-\lambda\widetilde{\jvet}
\]

\noindent where 
\[
\widetilde{\jvet}=\frac{1}{2}\left.\nabla g(\zvet)\right|_{\zvet=\ytilde}
\]

\begin{figure}
	\centering
	\includegraphics[trim=1cm 1cm 1cm 1cm, clip, width=0.45\textwidth]{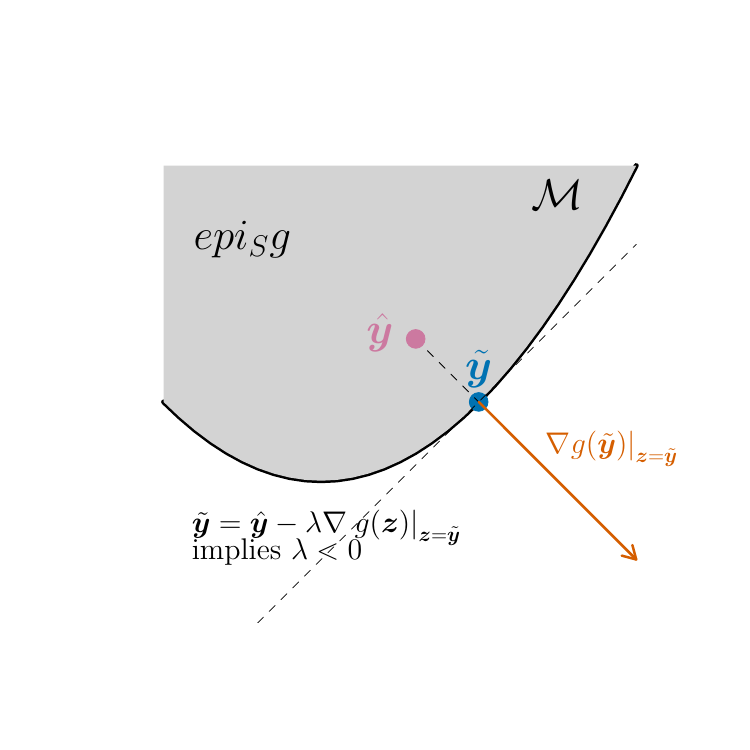}
	\includegraphics[trim=1cm 1cm 1cm 1cm, clip, width=0.45\textwidth]{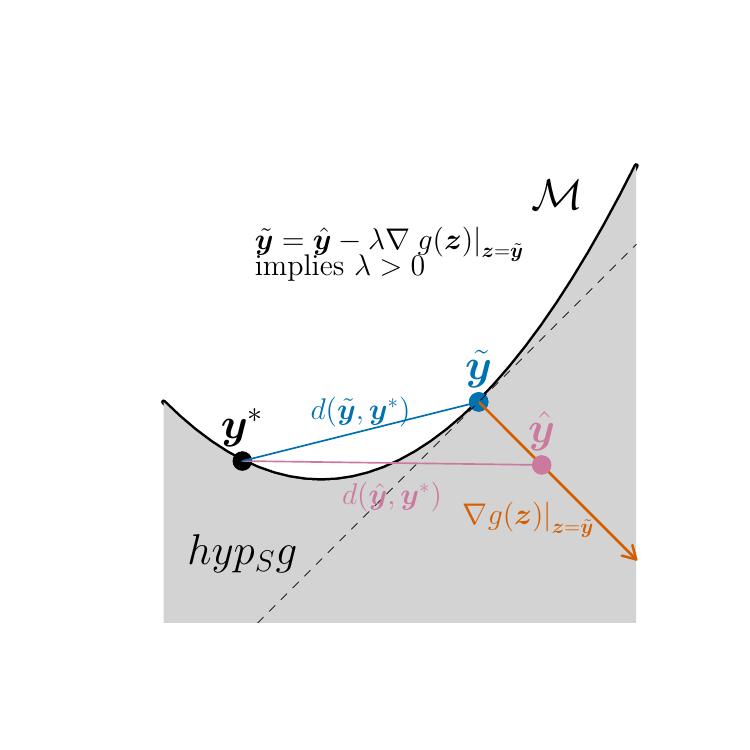}
	\caption{Schematic demonstrating non linear constrained reconciliation in the one constraint case. On the right panel is the case where the base forecast ($\widehat{\yvet}$, shown in pink) lies in the strict epigraph of the constraint $epi_S g(\cdot)$, shown as a grey shaded area. The reconciled forecast $\ytilde$ shown in blue is obtained by `projecting' along the gradient vector at $\ytilde$. This gradient vector denoted $\nabla g(\widetilde{\yvet})$ is depicted as a orange arrow starting from $\widetilde{\yvet}$. In this case $\lambda<0$. A hyperplane that is orthogonal to $\nabla g(\widetilde{\yvet})$ is shown as a dotted line. On the right panel the same information is shown when the base forecast lies in the strict hypograph, $hyp_S g(\cdot)$. Also shown on the right panel is an arbitrary point $\ystar$ and the triangle formed between $\yhat$, $\ystar$ and $\ytilde$. Critical to the proof of \autoref{lemma:lemma2} is that the angle corresponding to the vertex at $\ytilde$ is greater than $\pi/2$ radians.}
	\label{fig:epihypo}
\end{figure}

It is instructive to consider three sets of points. First, the set $\yvet:g(\yvet)=0$ (already denoted as $\cM$), second, the set $epi_S g:=\yvet:g(\yvet)<0$ a convex set known as the \textit{strict epigraph} and third, the set $hyp_S g:=\yvet:g(\yvet)>0$ or \textit{strict hypograph}. Since $g(\yvet)$ is positive for $\yvet\in hyp_S  g$, the gradient vector $\widetilde{\jvet}$ evaluated at any point in $\cM$ `points' towards the strict hypograph, i.e. $\ytilde+\epsilon \widetilde{\jvet}\in hyp_S g$ for arbitrarily small $\epsilon>0$. Therefore, for values of $\yhat\in hyp_S g$, $\lambda$ will be positive, while for values of $\yhat\in epi_S g$, $\lambda$ is negative. This is depicted in \autoref{fig:epihypo}.

We first prove that reconciliation is guaranteed to improve forecast accuracy in the one constraint case when $\yvet\in epi_S g$, before generalising this result to the case with multiple constraints. The proof relies on the following intermediate result.

\begin{lemma}\label{lemma:lemma1}
Let $\ystar$ be an arbitrary point such that $\ystar\in\cM$. Then 
$$
\langle(\ystar-\ytilde),\widetilde{\jvet}\rangle\leq 0
$$
\end{lemma}
\begin{proof}
	By the supporting hyperplane theorem, there exists a hyperplane that is tangent to the graph $g(\yvet)=0$ at $\ytilde$ with all points $\yvet:g(\yvet)=0$ lying in one halfspace of the hyperplane (or for linear $g$, on the hyperplane). This hyperplane is orthogonal to $\widetilde{\jvet}$ with $\widetilde{\jvet}$ lying in the opposite halfspace to the graph of $g(\mathbf{y})=\mathbf{0}$. Let $\theta$ be the angle formed between $(\ystar-\ytilde)$ and $\widetilde{\jvet}$. Since the angle between $\widetilde{\jvet}$ and the supporting hyperplane is $\pi/2$ and since $(\ystar-\ytilde)$ is on the opposite side of the supporting hyperplane (or in the limiting case, on the supporting hyperplane), then $\theta$ must be an obtuse angle, i.e, $\theta\geq\pi/2$. Therefore
	\begin{align*}
		\cos(\langle(\ystar-\ytilde),\widetilde{\jvet}\rangle)&\geq\pi/2\\
		\Rightarrow\langle(\ystar-\ytilde),\widetilde{\jvet}\rangle& \geq arccos(\pi/2)\\
		\Rightarrow\langle(\ystar-\ytilde),\widetilde{\jvet}\rangle& \leq 0
	\end{align*}
\end{proof}

\begin{lemma}\label{lemma:lemma2}
	For a base forecast $\yhat\in hyp_S g$, reconciliation always improves the forecast in the sense that $d(\ytilde,\yvet)-d(\yhat,\yvet)<0\,\forall \yvet\in\cM$ . 
	\end{lemma}

\begin{proof}
By \autoref{lemma:lemma1}, for any $\ystar\in\cM$ $\langle(\ystar-\ytilde),\widetilde{\jvet}\rangle \leq 0$. For the case of equality, $(\ystar-\ytilde)$ are at right angles $(\yhat-\ytilde)$, and the result stated in \autoref{lemma:lemma2} is a simple consequence of Pythagoras' theorem. For the case of strict inequality
$$
sign\left(\langle(\ystar-\ytilde),\lambda\widetilde{\jvet}\rangle\right)=-sign\left(\lambda\right)
$$
which after rearranging \autoref{eq:reco_sol} and substituting implies that 
\begin{equation}
	sign\left(\langle(\ystar-\ytilde),(\yhat-\ytilde)\rangle\right)=-sign\left(\lambda\right)\label{eq:ip}
\end{equation}
\end{proof}

Now consider the triangle formed by $\yhat$, $\ytilde$ and $\ystar$. This is depicted on the right panel of \autoref{fig:epihypo} for the two variable case, however, the following argument generalises to higher dimensions. The angle between the line from $\yhat$ to $\ytilde$ and the line from $\ystar$ and $\ytilde$ is given by taking the cosine of the inner product in \autoref{eq:ip}. When $\yhat\in hyp_S g$, $\lambda$ is positive, the inner product is negative and the angle is greater than or equal to $\pi/2$. In this case, the line opposing this angle, i.e. the line from $\yhat$ to $\ystar$ with length $d(\yhat,\ystar)$ will be the longest line in the triangle. It will be longer that the line from $\ytilde$ to $\ystar$ with length $d(\ytilde,\ystar)$. Therefore $d(\ytilde,\ystar)-d(\yhat,\ystar)$, thus completing the proof.

Now consider the case where there are multiple constraints as outlined in \autoref{eq:reco_sol} leading to $C$ columns in $\widetilde{\Jvet}$ given by $\widetilde{\jvet}_1,\dots,\widetilde{\jvet}_C$ and $C$ Lagrange multipliers $\lambda_1,\dots,\lambda_C$. By \autoref{lemma:lemma1}, $\langle(\ystar-\ytilde),\widetilde{\jvet}_c\rangle\leq 0$ for all $c$. 

\begin{theorem}\label{thm:thm2}
For the case with multiple constraints, $d(\ytilde,\yvet)-d(\yhat,\yvet)<0\,\forall \yvet\in\cM$ is guaranteed whenever $\yhat\in\bigcap\limits_{c=1}^{C}hyp_S g_c$, i.e. in the intersection of hypographs of all constraints.
\end{theorem}
\begin{proof}
	Following similar reasoning to the proof from the previous section
	\begin{align*}
		\langle(\ystar-\ytilde),(\yhat-\ytilde)\rangle&=\langle(\ytilde-\ystar),\sum\limits_{c=1}^{C}\lambda_c\widetilde{\jvet}_c\rangle\\
		&=\sum\limits_{c=1}^{C}\lambda_c\langle(\ystar-\ytilde),\widetilde{\jvet}_c\rangle\\
	\end{align*}
All inner products in the sum on the previous line are negative (or zero). $d(\ytilde,\ystar)<d(\yhat,\ystar)\,\forall \ystar\in\cM$ is guaranteed when the entire sum is negative. This can only be guaranteed for all $\ystar$, when all $\lambda_k$ are positive. This occurs when $\yhat\in\bigcap\limits_{c=1}^{C}hyp_S g_c$.
\end{proof}

For the case where $\yhat$ lies in an intersection of hypographs and epigraphs the results in Section~\ref{sec:guar} continue to hold. We can see how this influences the expression for the radius of the ball in \ref{eq:radius}. This is most clear in the two constraint case. Focusing attention on the final term in the expression for the radius, $\frac{\mu^2}{4}\lambdavet\widetilde{J}'\widetilde{J}\lambdavet$. We can then consider two cases. The first is where the base forecast lies in the epigraph of one constraint and in the hypograph of the other in which case $\lambda_1$ and $\lambda_2$ have opposite signs. In this case, the radius will be larger (reconciliation is better) when $\langle\widetilde{\jvet}_1,\widetilde{\jvet}_2\rangle$ is lower. This occurs when the gradient vectors at the reconciled forecast are very different across the two constraints. Alternatively, if the base forecast is in both epigraphs ($\lambda_1$ and $\lambda_2$ have the same sign), the radius will be larger when $\langle\widetilde{\jvet}_1,\widetilde{\jvet}_2\rangle$ is larger. This corresponds to gradients evaluated at the reconciled forecast that are similar for the two constraints.

\subsection{Discussion of theoretical results}\label{sec:the_disc}

In summary, forecast reconciliation will lead to improvements in forecast accuracy when:

\begin{itemize}[nosep]
	\item The constraint functions are convex, or failing that, when the probability that $\Yvet\in\mathcal{B}$ is large, where  i.e. $\mathcal{B}$ is a ball in which convexity holds.
	\item The probability that $\yhat$ is the intersection of hypographs is large.
	\item The probability that $\Yvet\in\mathcal{B}_{\ytilde,r}\cap\cM$ is large, as occurs when $\ytilde$ is close to a high probability region of  $\Yvet$.
	\item The radius of $\mathcal{B}_{\ytilde,r}\cap\cM$ is large, which given all else stays constant can occur when
	\begin{itemize}[nosep]
		\item The base and reconciled forecast are far apart
		\item The constraint functions have lower curvature
		\item The gradients of different constraint functions are similar if the base forecast is in the epigraph of constraints (and the opposite holds for the case where the base forecast is in a mixture of hypographs and epigraphs).
	\end{itemize}
\end{itemize}

Note that these theoretical results also motivate the use of a weighting matrix $\bm{W}$ in the objective function. Ideally, the space should be transformed so that larger forecasting errors occur in a direction orthogonal to constraints ($\ytilde$ is far from $\yhat$), while forecast errors along the constraints are small ($\ytilde$ is close to the true mean, so that the ball around $\ytilde$ is a high probability region).
Furthermore, the theoretical results can be informative if some a priori information is available about the bias of forecasts. It should be noted that the optimisation procedure outlined above is a non-linear mapping of $\yhat$ to $\ytilde$. As such, it is not possible to prove that unbiased base forecasts remain unbiased after reconciliation, as is the case for purely linear constraints. However, in a given application, it may be the case that there is some systematic bias that base forecasts are more likely to be in the hypograph of the constraints. This scenario would be favourable to reconciliation methods, while the opposite (a systematic bias towards the epigraph) would suggest that it is less likely that reconciliation improves forecast accuracy.

\section{Implementation details} \label{sec:alg}

We developed an \textsf{R} package \pkgname(Non-Linearly Constrained forecast Reconciliation\footnote{GitHub: \href{https://github.com/danigiro/nlcReco}{\texttt{danigiro/nlcReco}}}), to address the challenge of forecast reconciliation under non-linear constraints using the \texttt{FoReco} package \citep{FoReco} to perform standard computation for linear reconciliation. In the empirical applications, we considered three different covariance matrix in \autoref{eq:objc} that have been proposed in the forecast reconciliation literature with linear constraints:
\begin{itemize}[nosep]
	\item \textbf{ols} (\textit{ordinary least squares} approach): $\widehat{\Wvet}_{ols}=\Ivet_n$ \citep{hyndman2011} 
	\item \textbf{wls} (\textit{weighted least squares} approach): $\widehat{\Wvet}_{wls} = \Ivet_{n} \odot \widehat{\Wvet} $ \citep{hyndman2016}
	\item \textbf{shr} (\textit{shrinkage} approach): $\widehat{\Wvet}_{shr}=\widehat{\lambda} \widehat{\Wvet}_{wls} +(1-$ $\widehat{\lambda}) \widehat{\Wvet}$ \citep{wickramasuriya2019},
\end{itemize}
where $\widehat{\Wvet}=\frac{1}{T} \sum_{t=1}^{T} \widehat{\evet}_{t} \widehat{\evet}_{t}^{\prime}$ is the covariance matrix of the one-step ahead in-sample forecast errors $\widehat{\mathbf{e}}_{t}$ \citep{wickramasuriya2019}, the symbol $\odot$ denotes the Hadamard product, and $\widehat{\lambda}$ is an estimated shrinkage coefficient \citep{ledoit2004a}.

The reconciliation algorithm in \pkgname employs a sequential quadratic programming (SQP) approach, which is well-suited for non-linearly constrained gradient-based optimization. This method supports both inequality and equality constraints, making it versatile for a range of practical applications (i.e. non-negative forecast reconciliation). The SQP algorithm is detailed in  \cite{Kraft1988-uj, Kraft1994-wz}. The implementation of this algorithm is provided by the \textsf{R} package \texttt{nloptr} \citep{NLopt}, and a \textsf{Python} version is also available through the \texttt{SciPy} library \citep{SciPy-NMeth}. 

\section{Simulations}\label{sec:sim}

In this section, we investigate two simulated examples. The first is a simple bivariate case with one constraint given by a quartic equation. For this example the Gaussian curvature has a form that, while simple, varies over the manifold so as to demonstrate the role of curvature as explained in Section~\ref{sec:the_disc}. Since this constraint is also an example of a convex function, it also serves to demonstrate the theoretical results discussed in Section~\ref{sec:convex}. The second simulation setting is neither convex nor concave and refers to the case of a ratio. This constraint is commonly seen in practice, including in our empirical studies in Section~\ref{sec:app}.

\subsection{Simulation 1: Constraint is quartic (convex)}

\begin{figure}[!t]
	\centering
	\includegraphics[width=0.65\textwidth]{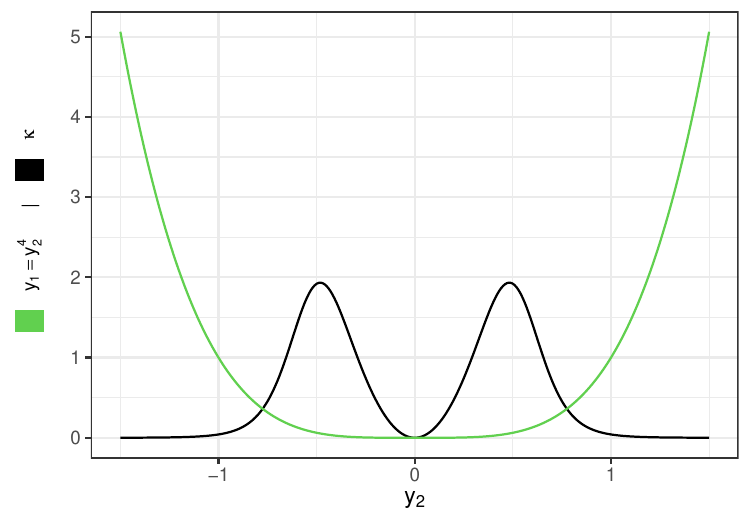}
	\caption{Illustration of the constraint $y_1 = y_2^4$ shown in green, with the associated Gaussian curvature $\kappa(y_2)$ computed along the curve and represented as a function of $y_2$ in black.}
	\label{fig:sim_convex_constraint}
\end{figure}

The first constraint we consider is $g(\yvet) = y_1 - y_2^4$.  By rewriting the constraint as $y_1 = f(y_2) = y_2^4$.  the Gaussian curvature, following \cite{doCarmo2016}, is given by
\[
\kappa(y_2) = \frac{|f''(y_2)|}{[1 + (f'(y_2))^2]^2} = \frac{12 y_2^2}{(1 + 16 y_2^6)^2}.
\]
As shown in \autoref{fig:sim_convex_constraint}, the curvature tends towards zero at $y_2 = 0$ and as $y_2 \rightarrow -\infty,\infty$, while it increases sharply over the interval $0 < |y_2| < 1$, reaching a maximum near $|y_2| = 0.5$. This indicates the presence of two highly curved zones in the region $0 < |y_2| < 1$ and flatter areas  around the origin and towards extreme values of $y_1$.

The true value (i.e., the target of our forecast), is set to $\yvet =  \begin{bmatrix}y_1 = y_2^4, & y_2\end{bmatrix}'$, with  $y_2$ set to the values on an evenly spaced grid between -1.5 and 1.5 with increments of 0.01. Base forecasts are generated according to 
$$
\widehat{\yvet} = \begin{bmatrix}
	\widehat{y}_1 \\
	\widehat{y}_2
\end{bmatrix} \sim \mathcal{N}_2\left(\widehat{\bm{\mu}}, \Sigmavet\right), \quad \text{where} \quad 
$$
where $\Sigmavet = \mathrm{diag}\left([0.1\; 0.1]\right)$, and the mean vector is given by
$$
\widehat{\bm{\mu}} = \begin{bmatrix}
	m(\widehat{y}_2 - y_2) + y_2^4 \\
	y_2 - \displaystyle\frac{\beta}{\sqrt{m^2+1}}
\end{bmatrix},
\quad \text{with} \quad 
m = \left(\left.\frac{\partial g(\yvet)}{\partial y_2}\right|_{\yvet}\right)^{-1}.
$$

\begin{figure}[!t]
	\centering
	\includegraphics[width=0.55\textwidth]{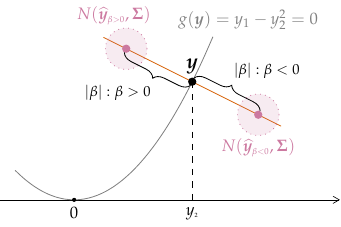}
	\caption{Illustration of the data-generating process (DGP) for the base forecasts under the constraint $y_1 = y_2^4$. The true values lie on the grey curve, while the base forecasts are generated around a mean shifted along the direction orthogonal to the tangent at the true point, controlled by the distortion parameter $\beta$.}
	
	\label{fig:sim_convex_new_DGP}
\end{figure}

Note that a single parameter controls the behavior of base forecasts in the simulation. The value of $\beta$ shifts the location of base forecasts in an orthogonal direction \textit{away} from the constraint surface. When $\beta = 0$, base forecasts are centered on $\yvet$; for $\beta > 0$, base forecasts are biased towards the epigraph, while for $\beta < 0$ they are biased towards the hypograph. This is depicted in \autoref{fig:sim_convex_new_DGP}.

For each true value, 1000 base forecasts are simulated according to the process described above.  Each of these base forecasts is reconciled using the OLS approach detailed in \autoref{sec:alg} to obtain the reconciled values $\widetilde{\yvet}$. To evaluate performance, we compare the base and reconciled forecasts with respect to their Euclidean distance from the true value, defined as $d(\widehat{\yvet}) = \lVert \widehat{\yvet} - \yvet \rVert_2$ and $d(\widetilde{\yvet}) = \lVert \widetilde{\yvet} - \yvet \rVert_2$, respectively. For each true value, we compute the proportion of simulations in which $d(\widehat{\yvet}) > d(\widetilde{\yvet})$, indicating how often the reconciled forecast is more accurate than the base forecast.

\begin{figure}[!t]
	\centering
	\includegraphics[width=\textwidth]{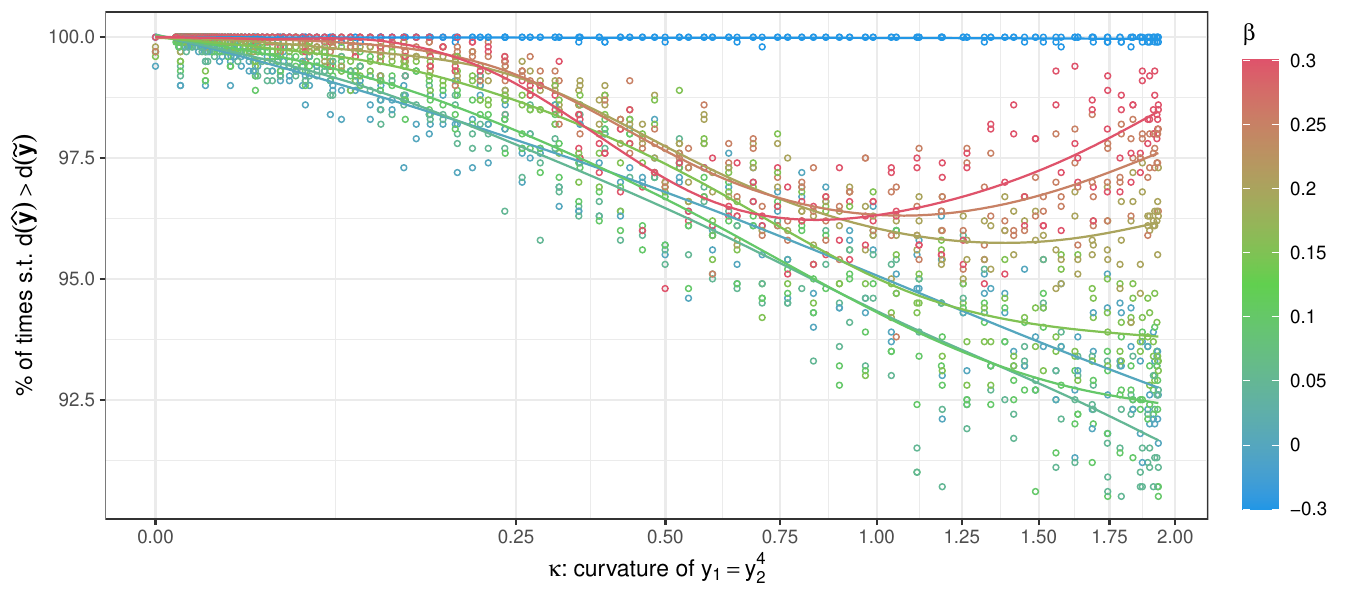}
	\caption{Results for simulation based on constraint $y_1=y_2^4$. Each point refers to a single target value, for which 1000 base forecasts are generated and reconciled. The y-axis depicts the proportion of times that reconciled forecasts outperform base forecasts. The x-axis depicts the corresponding Gaussian curvature. Different colors indicate different levels of bias controlled by the parameter $\beta$.}
	\label{fig:sim_convex_res}
\end{figure}

\autoref{fig:sim_convex_res} shows these percentages as a function of both the distortion parameter $\beta$ and the constraint curvature $\kappa$, with smoothed curves included for visual clarity. When $\beta = -0.3$ (when the base forecasts are biased towards the hypograph), reconciled forecasts leads to improvements over base forecasts with probability near 1, regardless of the curvature. This can be explained by Lemma~\ref{lemma:lemma2}, projecting toward the constraint surface from the hypograph moves the forecast closer to the truth. For $0 \leq \beta \leq 0.15$, a markedly different behavior emerges: the percentage of cases where reconciled forecasts outperform base forecasts decreases with curvature. In high-curvature regions, reconciliation has a lower (albeit still high) probability of being more accurate than the base forecast. This demonstrates the role of the inner products of Jacobian terms in Equation~\ref{eq:radius}.  For high levels of bias ($\beta \geq 0.15$), the probability that reconciliation improves base forecasts begins to increase at higher levels of curvature (e.g. for $\kappa > 0.75$). Although this would seem to contradict our theoretical results, it should be noted that these arguments rely on local approximations that can break down when the base forecasts are further away from the constraint surface and the global properties of the constraint surface become relevant. It is nonetheless encouraging that in these scenarios, forecasts with large initial incoherencies benefit from reconciliation.

\subsection{Simulation 2: Constraint is a ratio (neither convex nor concave)}

\begin{figure}[!t]
	\centering
	\includegraphics[width=0.45\textwidth]{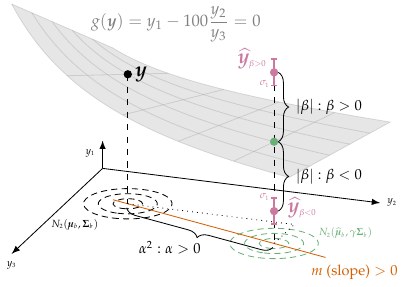}\hfill\vline\hfill
	\includegraphics[width=0.45\textwidth]{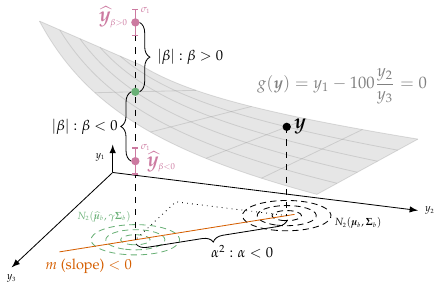}
	\caption{Visual data-generating process (DGP) for true values and base forecasts using the convex function $g(\yvet) = y_1 - 100\displaystyle\frac{y_2}{y_3}$. On the left, the parameter $\alpha \geq 0$, and on the right $\alpha < 0$.}
	\label{fig:sim_noconvex}
\end{figure}

The second simulation considers the function $g(\yvet) = y_1 - 100\displaystyle\frac{y_2}{y_3}$ which is neither convex nor concave. \autoref{fig:sim_noconvex} shows the mechanism used to simulate both true values and base forecasts. For the true values, we generate $y_2$ and $y_3$ from a multivariate normal distribution with mean $\muvet_b = [100 \; 300 ]'$ and variance covariance matrix $\Sigmavet_b$ with diagonal elements set to 5 and 10 and a correlation set to $\rho \in \{-0.8,\; -0.4,\; 0,\; 0.4,\; 0.8\}$. For each draw, the value of $y_1$ is determined via the constraint. Overall, 1000 such true values are generated for each value of $\rho$.

For each true value, base forecasts are obtained as
$$
\widehat{\yvet} = \begin{bmatrix}
	\widehat{y}_1 \\ \widehat{y}_2 \\ \widehat{y}_3
\end{bmatrix} = \begin{bmatrix}
	100\frac{\widehat{y}_2}{\widehat{y}_3} + \varepsilon \\ \widehat{y}_2 \\ \widehat{y}_3
\end{bmatrix}
$$
where $\varepsilon\sim N\left(\beta, 100\right)$ with $\beta \in \{-25,\; -10,\; 0,\; 10,\; 25\}$ and 
$$
\begin{bmatrix}
	\widehat{Y}_2 \\ \widehat{Y}_3
\end{bmatrix} \sim N_2\left(\widehat{\muvet}_b = \muvet_b + \deltavet,
\gamma\Sigmavet_b\right)
$$
with $\gamma \in \{0.5,\; 1,\; 1.5\}$,
$$
\deltavet =
\begin{bmatrix}
1 \\
\tan(m)
\end{bmatrix} \frac{\alpha}{\sqrt{\tan^2(m)+1}},
$$
and $m \in \{-\pi/4,\; 0\}$, $\alpha \in \{-50,\; -25,\; 0,\; 25,\; 50\}$. Overall five parameters control the simulation:
\begin{itemize}[nosep]
\item $\beta$ biases the base forecast away from the coherent manifold in the direction of $y_1$. For $\beta=0$ the base forecast has an expected value that is coherent.
\item $m$ controls the direction of the bias along the coherent manifold. When $m = 0$, the location of base forecasts is displaced from the true value entirely along the $Y_2$-axis. In this case, the induced change in $Y_1$ is linear with no curvature. When $m = -\pi/4$, the base forecasts are displaced along the line $Y_3=-Y_2$. Along this direction the constraint has high curvature, small biases in $(y_2,y_3)$ translate into disproportionately large distortions in $y_1$.
\item $\alpha$ controls the magnitude of bias along this coherent manifold. A positive $\alpha$ moves the base forecast mean further along the direction determined by $m$, while a negative $\alpha$ shifts it in the opposite direction.
\item $\gamma$ controls the variability of base forecasts with larger values of $\gamma$ leading to more extreme values of the base forecast.
\item $\rho$ controls dependence between bottom level true values and (and also the corresponding base forecasts).
\end{itemize}

\begin{figure}[!t]
	\centering
	\includegraphics[width=\textwidth]{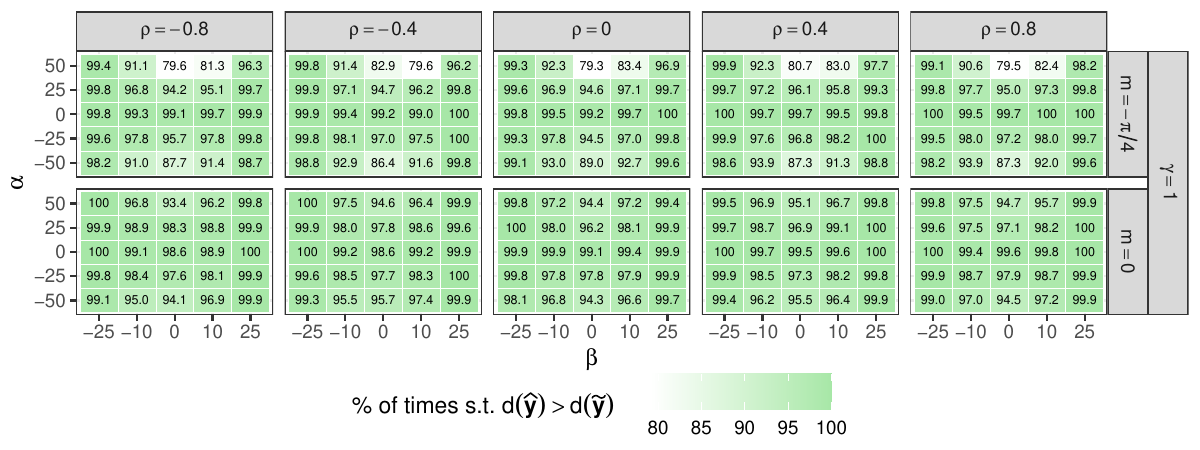}
	\caption{Subset of the simulation results showing the proportion of times that reconciled forecasts are closer to the target value than base forecasts. Results are shown for different simulation parameters.}
	\label{fig:res_sim2}
\end{figure}

We reconcile each base forecast to derive the reconciled forecast $\widetilde{\yvet}$ using the OLS approach proposed in \autoref{sec:alg}. As a measure of forecast accuracy, we consider the distance of each base forecast ($\widehat{\yvet}$) and each reconciled forecast ($\widetilde{\yvet}$) from the true value ($\yvet$). A subset of the results is shown in \autoref{fig:res_sim2}, while the complete figures for each parameter considered are available in the online supplementary material. \autoref{fig:res_sim2} displays the proportion of values for which reconciled forecasts are more accurate than base forecasts with darker shading indicating more accurate base forecasts. Generally, we observed an improvement of over 85\% in areas where the bias is in directions of lower curvature ($m=0$). The only cases where the proportion is below 85\% occurs when base forecasts there is no (or little) bias away from the coherent manifold ($\beta \in \{0, 10\}$), but where base forecasts are biased along the coherent manifold in a direction of high curvature (with $\alpha = 50$ and $m = -\pi/4$). This is line with our theoretical results, in general, reconciliation has a high likelihood of improving forecast accuracy, except in situations where the constraint is highly curved, and incoherency is not too severe. It is nonetheless noting that even in this worst case scenario, the probability that reconciliation improves forecast accuracy remains above 79\%.

\section{Empirical applications}\label{sec:app}

\subsection{Mortality rates with a simple hierarchy}\label{sec:Mrates}

\subsubsection{Set up}

As a first empirical application, we consider the annual male U.S. mortality data spanning the period 1969--2019. Death counts are obtained from two complementary sources: the National Center for Health Statistics (NCHS) for the years 1969--2004 and the CDC WONDER online database for the years 2005--2019. Corresponding exposure data are drawn from the Surveillance, Epidemiology, and End Results (SEER) program, which provides estimates of the annual population at risk by single year of age up to 85+. In order to avoid distortions due to the COVID-19 pandemic, the sample is truncated at 2019 as described in \cite{Li2024}. 

\begin{figure}[!t]
	\centering
	\begin{tikzpicture}[baseline=(current  bounding  box.center),
every node/.append style={shape=rectangle, anchor = center,
	draw=black},
minimum width=1.5cm]

\foreach[count=\i, evaluate=\i as \x using ((\i-1)*1.75)] \y in {NE,MA,ENC,WNC,SA,ESC,WSC,MT,PA}
{
    \node[color=black] at (\x, 0) (b\i){\y};
}

\node[color=black] at ({((9-1)*1.75)/2}, 1.25) (u1){USA};

\foreach \i in {1,...,9} {
	\relation{0.25}{b\i}{u1};
}
\end{tikzpicture}
	\caption{Hierarchical structure of the U.S. mortality dataset, with the national total at the top level and nine census divisions at the second level, following the classification of the United States Census Bureau.}
	\label{fig:mrate_hier}
\end{figure}
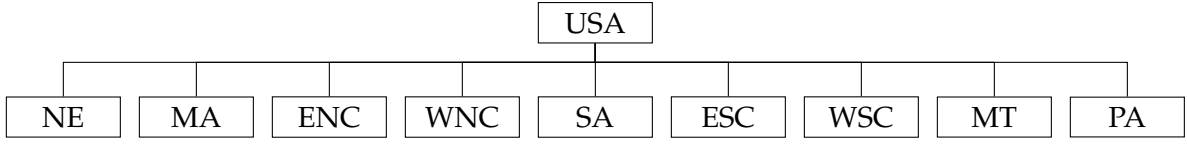

Our analysis focuses on three series of interest: death counts $(D)$, population exposures $(P)$, and mortality rates $(R)$, defined as  
\begin{equation}\label{eq:Mrates}
	R = \frac{D}{P}.
\end{equation}
We adopt the hierarchical structure for the population exposures and death counts used by the United States Census Bureau and aggregate data at the level of census divisions. This results in a two-level hierarchy as shown in  \autoref{fig:mrate_hier}, with the national total at the top and nine census divisions at the bottom, namely: New England (NE), Middle Atlantic (MA), East North Central (ENC), West North Central (WNC), South Atlantic (SA), East South Central (ESC), West South Central (WSC), Mountain (MT), and Pacific (PA). 
For the underlying hierarchical structure, we have the following aggregation constraints: 
\begin{align}
D_{USA}=\sum_{i\in{regions}} D_i, \label{equ:17}\\ 
P_{USA}=\sum_{i\in{regions}} P_i. \label{equ:18}
\end{align}

Our proposed reconciliation approaches in \autoref{sec:alg} simultaneously incorporate both the linear aggregation constraints imposed by the hierarchy as stated in \autoref{equ:17} and \autoref{equ:18}, as well as the non-linear relationship in \autoref{eq:Mrates} between the variables. Our goal is to generate coherent forecasts for all three time series, with particular emphasis on the mortality rate $R$. To evaluate the performance of our proposed non-linear constrained reconciliation, we include two other benchmark approaches:
\begin{itemize}[nosep, leftmargin = 1.5cm]
	\item[\textbf{bu}:] The bottom up method. The forecasts of the bottom-level time series (Death counts and Population exposures for the 9 census divisions) are aggregated to obtain the national total; then Equation \ref{eq:Mrates} is used to obtain the mortality rates' forecasts.
	\item[\textbf{LH}:] {The approach proposed by \cite{Li2021}.This approach applies the MinT reconciliation \cite{wickramasuriya2019} using a modified summation matrix to reformulate the aggregation constraint as a linear problem. Note that this method does not produce reconciled forecasts for $D$ and $P$. }
\end{itemize}

To evaluate and compare the forecasting performance of different methods, we perform an expanding window forecast experiment. The initial training sample covers the period 1969--1996, after which the models are re-estimated at each forecast origin, producing $L_{10}=13$ sets of forecasts with horizon $h=10$, $L_{9}=14$ sets with $h=9$, ..., $L_{2}=21$ sets with $h=2$ and $L_{1}=22$ sets with $h=1$. The base forecasts are obtained by fitting the Lee–Carter model \citep{Lee1992} to the mortality rates, while the base forecasts of death counts and population exposures are generated using ARIMA models \citep{BoxJenkins1976}. ARIMA orders are selected using the ``\texttt{auto.arima}'' function in the \texttt{forecast} package in \textsf{R} \citep{hyndman2008a,Rforecast}. 

Let $y_{i,j,h,l}$, $\widehat{y}_{i,j,h,l}$, and $e_{i,j,h,l} = y_{i,j,h,l} - \widehat{y}_{i,j,h,l}$ denote, respectively, the observed value, the base forecast, and the corresponding forecast error for variable $i \in \{R, D, P\}$, geographical unit $j \in \{\text{USA}, \text{NED}, \ldots, \text{PD}\}$, forecast horizon $h = 1, \ldots, 10$, and $l = 1, \ldots, L_h$. To evaluate forecast accuracy, we employ root mean squared error (RMSE) as error measures:  
\begin{equation}
	RMSE_{i,j,1:H}^{comb} = \sqrt{
		\frac{1}{H}\sum_{h=1}^H \sum_{l=1}^{L_h} \frac{1}{L_h} e_{i,j,h,l}^2
		}.
\end{equation}

To summarise performance across census divisions and variables, we compute geometric means of accuracy ratios relative to the base forecasts:  
\begin{equation}
	gmRMSE_{1:H}^{comb} = \left(\prod_{i,j}\frac{RMSE_{i,j,1:H}^{comb}}{RMSE_{i,j,1:H}^{base}} \right)^{\frac{1}{30}}.
\end{equation}

Finally, in order to assess whether observed differences in predictive accuracy are statistically significant, we complement these accuracy measures with a number of widely used tests. Specifically, we apply the \cite{DieboldMariano1995} test, the Model Confidence Set procedure of \citet{Hansen2011}, and the post hoc multiple comparison with the best (MCB) Nemenyi test \citep{koning2005, kourentzes2019, makridakis2022}, which together provide a comprehensive framework for evaluating relative forecast performance.  

\subsubsection{Results}

We present the empirical results of applying the forecast reconciliation methods and benchmarks to the male U.S. mortality dataset, focusing on the squared loss function $gmRMSE$ as the primary measure of forecast accuracy.

\begin{table}[h!]
	\centering
	\begingroup
	\spacingset{1.1}
	\small
	
\begin{tabular}[t]{r|ccc|ccc|ccc}
\toprule
\multicolumn{1}{c}{\textbf{}} & \multicolumn{3}{c}{\textbf{USA}} & \multicolumn{3}{c}{\textbf{C.D.}} & \multicolumn{3}{c}{\textbf{All}} \\
\textbf{App.} & Rates & Others & All & Rates & Others & All & Rates & Others & All\\
\midrule
\addlinespace[0.3em]
\multicolumn{1}{c}{} & \multicolumn{9}{c}{$h=1$}\\
bu & 0.408 & 1.033 & 0.758 & 0.510 & 1.000 & 0.799 & 0.498 & 1.003 & 0.795\\
LH & 1.015 &  &  & 0.997 &  &  & 0.999 &  \vphantom{1} & \\
ols & 0.375 & \em{0.997} & 0.720 & 0.491 & 0.990 & 0.784 & 0.478 & 0.991 & 0.777\\
wls & \textbf{0.374} & \textbf{0.995} & \textbf{0.718} & \em{0.490} & \textbf{0.989} & \em{0.783} & \em{0.477} & \textbf{0.990} & \textbf{0.776}\\
shr & \em{0.375} & 0.997 & \em{0.720} & \textbf{0.490} & \em{0.989} & \textbf{0.782} & \textbf{0.477} & \em{0.990} & \em{0.776}\\
\addlinespace[0.3em]
\multicolumn{1}{c}{} & \multicolumn{9}{c}{$h=5$}\\
bu & 0.747 & 1.033 & 0.927 & 0.806 & 1.000 & 0.931 & 0.800 & 1.003 & 0.930\\
LH & 1.015 &  &  & 0.997 &  &  & 0.999 &  & \\
ols & 0.695 & \textbf{0.999} & 0.885 & 0.769 & 0.982 & 0.905 & 0.761 & 0.983 & 0.903\\
wls & \em{0.694} & \em{1.000} & \em{0.885} & \em{0.759} & \em{0.978} & \em{0.899} & \em{0.752} & \em{0.980} & \em{0.897}\\
shr & \textbf{0.688} & 1.000 & \textbf{0.883} & \textbf{0.750} & \textbf{0.974} & \textbf{0.893} & \textbf{0.743} & \textbf{0.977} & \textbf{0.892}\\
\addlinespace[0.3em]
\multicolumn{1}{c}{} & \multicolumn{9}{c}{$h=10$}\\
bu & 0.956 & 1.033 & 1.007 & 1.008 & 1.000 & 1.003 & 1.003 & 1.003 & 1.003\\
LH & 1.014 &  &  & 0.997 &  &  & 0.999 &  & \\
ols & \em{0.892} & \textbf{1.000} & \textbf{0.962} & 0.963 & 0.985 & 0.978 & 0.956 & 0.986 & 0.976\\
wls & 0.892 & \em{1.004} & \em{0.965} & \em{0.943} & \em{0.980} & \em{0.967} & \em{0.938} & \em{0.982} & \em{0.967}\\
shr & \textbf{0.885} & 1.008 & 0.965 & \textbf{0.927} & \textbf{0.979} & \textbf{0.961} & \textbf{0.923} & \textbf{0.982} & \textbf{0.962}\\
\bottomrule
\end{tabular}

	\endgroup
	\caption{Mortality dataset. Average relative accuracy indexes ($gmRMSE$) where the reference forecasts are the base forecasts ($gmRMSE = 1$): lower values indicate better accuracy, bold highlights the best performance, italic indicates the second best.} 
	\label{tab:gmRMSE_mortality}
\end{table}

\autoref{tab:gmRMSE_mortality} reports the geometric mean of relative accuracy indexes ($gmRMSE$) for the different forecasting approaches, multiple forecast horizons, top-level (USA),  bottom-level (C.D.) and all series, and across the variable subsets (Rates, Others, and All). Across all horizons and subsets, the shr approach consistently achieves the lowest $gmRMSE$, indicating a substantial improvement over the base forecasts. The wls and ols reconciliations in \autoref{tab:gmRMSE_mortality} also perform well, often ranking as the second-best alternatives. The bu approach performs well for very short horizons, but its performance declines for longer-term forecasts, particularly for mortality rates. The LH approach has results very closed to the base. The improvements of shr is further reinforced by the Diebold-Mariano tests reported in the online supplementary material, which reveal statistically significant improvements over base forecasts across nearly all age groups and census divisions.

\begin{figure}[h!]
	\centering
    \includegraphics[width = \linewidth]{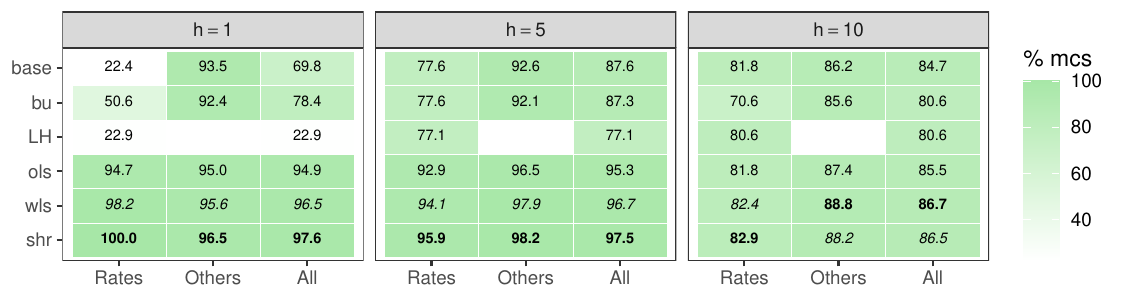}
    \caption{Mortality dataset. Model Confidence Set results at 95\% threshold using squared loss: percentage of times each forecasting approach is included in the best confidence model set for different forecast horizons ($h=1$, 5, and 10) and for the rates (Rates), the population exposures and death counts together (Others), and for all the variables (All). Values in bold indicate the method with the highest inclusion rate for a given metric and horizon. Higher percentages indicate more frequent inclusion in the best model set, reflecting greater relative forecast accuracy.}
	\label{tab:MCS_MSE_mortality}
\end{figure}

\begin{figure}[h!]
	\centering
	\includegraphics[width = 0.9\linewidth]{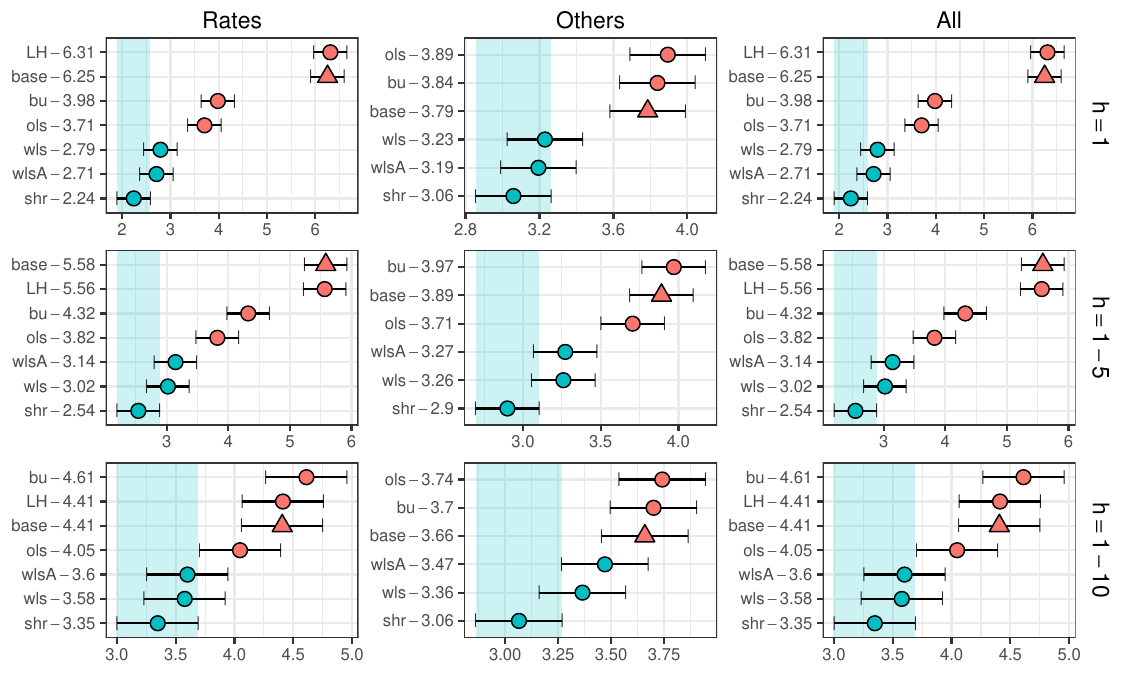}
	\caption{Mortality dataset. MCB Nemenyi test for the Australian electricity generation dataset using the RMSE at different forecast horizon ($h$ = 1, 5, and 10) for the rates (Rates), the population exposures and death counts together (Others) and all the variables (All). In each panel, the Friedman test $p$-value is reported in the lower-right corner. The mean rank of each approach is shown to the right of its name. Statistically significant differences in performance are indicated if the intervals of two forecast reconciliation procedures do not overlap. Thus, approaches that do not overlap with the green interval are considered significantly worse than the best, and vice versa.}
	\label{fig:MCB_MSE_mortality}
\end{figure}

\autoref{tab:MCS_MSE_mortality} presents the frequency with which each method is included in the best Model Confidence Set (MCS) for horizons $h=1$, 5, and 10. Once again, shr is almost always included, confirming its superior predictive accuracy. The wls and ols methods are frequently present as strong alternatives, while the base and bottom-up forecasts are less consistently selected, especially for longer horizons. Additional evidence is provided by 
Figure \ref{fig:MCB_MSE_mortality}, based on the MCB Nemenyi test, which further highlight that reconciliation methods accounting for non-linear constraints leads to statistically significant improvements in forecast accuracy.

Overall, the combined evidence from relative accuracy measures, \cite{DieboldMariano1995} tests, MCS and MCB demonstrates that forecast reconciliation, and in particular the shr approach, substantially improves the performance by jointly incorporating linear aggregation constraints and the non-linear dependence of mortality rates on deaths and exposures. The improvements are consistent across short-, medium-, and long-term horizons, illustrating the practical value of the proposed methodology for demographic forecasting applications.

\subsection{Unemployment rates with parallel hierarchies}

\subsubsection{Set up}
As a second empirical application, we turn to labour market data and consider monthly unemployment rates for Australia over the period January 1992 to April 2024 collected from the Australian Bureau of Statistics\footnote{\url{https://www.abs.gov.au/statistics/labour/employment-and-unemployment/labour-force-australia-detailed}}. This dataset provides a complementary illustration of the proposed methodology in a context that differs both in frequency and in economic interpretation from the mortality example. In particular, the unemployment rate is a key indicator for economic monitoring and policy analysis. 

The analysis focuses on four related variables: the unemployment rate $(R)$, the labour force $(T)$, the number of employed individuals $(E)$, and the number of unemployed individuals $(U)$. These variables are connected through the following equations:
\begin{equation}
	R = 100\frac{U}{T}, 
	\qquad 
	T = E + U.
\end{equation}

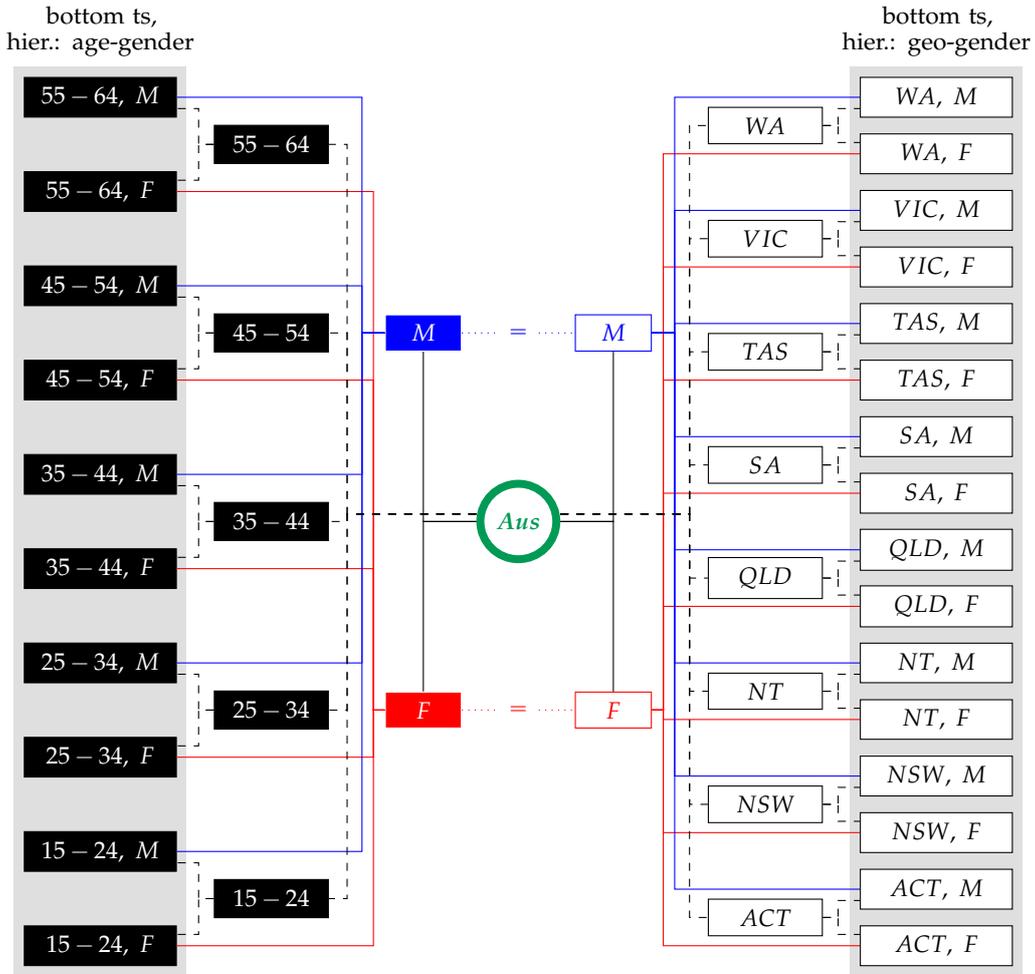
\begin{figure}[h!]
	\centering
	\begin{tikzpicture}[baseline=(current  bounding  box.center),
every node/.append style={shape=rectangle, anchor = center, font = \footnotesize,
	draw=black},
minimum width=1.5cm]

\foreach[count=\i, evaluate=\i as \x using ((\i-5.5)*1.25)] \y in {
{$15-24,\; F$},
{$15-24,\; M$}, 
{$25-34,\; F$}, 
{$25-34,\; M$}, 
{$35-44,\; F$},
{$35-44,\; M$},  
{$45-54,\; F$},  
{$45-54,\; M$},
{$55-64,\; F$}, 
{$55-64,\; M$}}
{
    \node[fill = black, text = white, minimum width=2cm] at (0, \x) (a\i){\y};
}

\foreach[count=\i, evaluate=\i as \x using ((\i-3)*2.5)] \y in {
{$15-24$},
{$25-34$},
{$35-44$},
{$45-54$},
{$55-64$}}
{
    \node[fill = black, text = white] at (2.25, \x) (ua\i){\y};
}

\foreach[count=\i, evaluate=\i as \x using ((\i-8.5)*0.75)] \y in {
{$ACT,\; F$}, 
{$ACT,\; M$}, 
{$NSW,\; F$}, 
{$NSW,\; M$}, 
{$NT,\; F$}, 
{$NT,\; M$}, 
{$QLD,\; F$}, 
{$QLD,\; M$},  
{$SA,\; F$}, 
{$SA,\; M$}, 
{$TAS,\; F$}, 
{$TAS,\; M$}, 
{$VIC,\; F$},  
{$VIC,\; M$}, 
{$WA,\; F$}, 
{$WA,\; M$}}
{
    \node[color=black, minimum width=2cm, fill = white] at (11, \x) (b\i){\y};
}

\foreach[count=\i, evaluate=\i as \x using ((\i-4.5)*1.5)] \y in {
{$ACT$},  
{$NSW$}, 
{$NT$}, 
{$QLD$}, 
{$SA$}, 
{$TAS$}, 
{$VIC$},  
{$WA$}}
{
    \node[color=black] at (8.75, \x) (ub\i){\y};
}

\node[circle, minimum width=1cm, color = ForestGreen, line width = 3pt] at (5.5, 0) (u1){$\bm{Aus}$};
\node[color=white, minimum width=1cm, fill = red, text = white] at (4.25, -2.5) (u2){$F$};
\node[color=white, minimum width=1cm, fill = blue, text = white] at (4.25, 2.5) (u3){$M$};

\node[color=black, minimum width=1cm, color = red] at (6.75, -2.5) (u21){$F$};
\node[color=black, minimum width=1cm, color = blue] at (6.75, 2.5) (u31){$M$};

\begin{pgfonlayer}{background}
   \node[fill = gray,draw = none, opacity = 0.25,fit=(a1) (a10), label = {[above, text width = 2.5cm, align = center]{bottom ts,\\[-0.15cm] hier.: age-gender}}] {};
   \node[fill = gray,draw = none, opacity = 0.25,fit=(b1) (b16), label = {[above, text width = 2.5cm, align = center]{bottom ts,\\[-0.15cm] hier.: geo-gender}}] {};
\end{pgfonlayer}

\draw[dotted, color = red] (u2) -- (u21) node[midway, fill = white, draw = none, minimum width=0cm] (eq1){$=$};

\draw[dotted, color = blue] (u3) -- (u31) node[midway, fill = white, draw = none, minimum width=0cm] (eq2){$=$};

\draw (u2.north) |- (u1.west);
\draw (u3.south) |- (u1.west);

\draw (u21.north) |- (u1.east);
\draw (u31.south) |- (u1.east);
	
\foreach \i in {1,3,...,9} {
	\draw[color = red] (u2.west) -- +(-0.15,0) |- (a\i.east);
}

\foreach \i in {2,4,...,10} {
	\draw[color = blue] (u3.west) -- +(-0.3,0) |- (a\i.east);
}

\foreach[count=\i, evaluate=\i as \x using int((\i+1)/2)] \y in {1,2,...,10}
{
  \draw[dashed] (ua\x.west) -- +(-0.2,0) |- ($(a\y.east)+(0, {iseven(\y)?-0.15:0.15})$);
}

\foreach \i in {1,...,5} {
	\draw[dashed] ($(u1.west)+(0, 0.1)$) -- +(-1.7,0) |- (ua\i.east);
}

\foreach \i in {1,3,...,15} {
	\draw[color = red] (u21.east) -- +(0.15,0) |- (b\i.west);
}

\foreach \i in {2,4,...,16} {
	\draw[color = blue] (u31.east) -- +(0.3,0) |- (b\i.west);
}

\foreach[count=\i, evaluate=\i as \x using int((\i+1)/2)] \y in {1,2,...,16}
{
  \draw[dashed] (ub\x.east) -- +(0.2,0) |- ($(b\y.west)+(0, {iseven(\y)?-0.15:0.15})$);
}

\foreach \i in {1,...,8} {
	\draw[dashed] ($(u1.east)+(0, 0.1)$) -- +(1.7,0) |- (ub\i.west);
}

\end{tikzpicture}	
	\caption{Hierarchical organisation of the Australian unemployment dataset. Both the employed and unemployed series are structured along parallel hierarchies, sharing the national (green) and gender (red and blue) aggregates, and further disaggregated by geography and age. This configuration results in a system of multiple time series subject to both linear aggregation constraints.}
	\label{fig:urate_hier}
\end{figure}

Both employment and unemployment series follow two hierarchies that share the national total (Aus) at the top level and the gender dimension at the first intermediate level. Additional intermediate disaggregations are given by geography and age groups, as illustrated in \autoref{fig:urate_hier}. As in the mortality application, this setting naturally combines linear aggregation constraints with a non-linear functional relationship, in this case arising from the definition of the unemployment rate. However, unlike standard grouped time series, here we are dealing with two hierarchies that partially overlap by sharing some upper-level series. This structure is more appropriately described as a system of linearly constrained multiple time series \citep{Girolimetto2024}, which represents a more general setting than classical hierarchical or grouped formulations proposed in \autoref{sec:Mrates}. For such a structure, a `bottom-up' method as applied in \autoref{sec:Mrates} is not possible.

To assess forecast accuracy, we design rolling origin experiments with fixed-length windows. The initial training sample spans January 1992 to December 2001, and subsequent windows are rolled forward month by month. This produces 245 forecast origins with horizon $h=24$, 246 origins with $h=23$, and so on, down to 267 origins with $h=2$ and 268 origins with $h=1$.  

For the base forecasts, we rely on ARIMA models \citep{BoxJenkins1976}, estimated automatically using the implementation provided by the \textsf{R} package \texttt{forecast} \citep{hyndman2008a,Rforecast}. Forecast evaluation follows the same framework as in the mortality application (see \autoref{sec:Mrates}), relying on RMSE as the loss function, geometric means of relative errors for comparison with benchmarks, and statistical tests such as the \cite{DieboldMariano1995} test, the Model Confidence Set \citep{Hansen2011}, and the post hoc multiple comparison with the best (MCB) Nemenyi test \citep{koning2005, kourentzes2019, makridakis2022}.

Overall, this second application illustrates how the proposed methodology can be effectively extended to economic time series, highlighting the flexibility of our approach in reconciling forecasts subject to both linear aggregation constraints and non-linear functional relationships, within a setting that is more complex in terms of both variables and constraints.

\subsubsection{Results}

\autoref{tab:gmRMSE_un} reports the geometric mean of relative accuracy indexes ($gmRMSE$) for the Australian unemployment system, computed for the upper ({Uts}), bottom ({Bts}), and all ({All}) series. Results are shown separately for the unemployment {rates}, for the remaining variables ({Others}), and for the entire hierarchy ({All}). Since the indexes are expressed relative to the base forecasts, the base values are always equal to one and are therefore omitted from the table.

\begin{table}[h!]
	\centering
	\begingroup
	\spacingset{1.1}
	\small
	
\begin{tabular}[t]{r|ccc|ccc|ccc}
\toprule
\multicolumn{1}{c}{\textbf{}} & \multicolumn{3}{c}{\textbf{Uts}} & \multicolumn{3}{c}{\textbf{Bts}} & \multicolumn{3}{c}{\textbf{All}} \\
\textbf{App.} & Rates & Others & All & Rates & Others & All & Rates & Others & All\\
\midrule
\addlinespace[0.3em]
\multicolumn{1}{c}{} & \multicolumn{9}{c}{$h=1$}\\
ols & 0.993 & 1.013 & 1.008 & 1.011 & 1.047 & 1.038 & 1.004 & 1.034 & 1.026\\
wls & \em{0.967} & \em{0.982} & \em{0.979} & \em{0.971} & \em{0.980} & \em{0.978} & \em{0.969} & \em{0.981} & \em{0.978}\\
shr & \textbf{0.945} & \textbf{0.967} & \textbf{0.961} & \textbf{0.962} & \textbf{0.976} & \textbf{0.972} & \textbf{0.955} & \textbf{0.972} & \textbf{0.968}\\
\addlinespace[0.3em]
\multicolumn{1}{c}{} & \multicolumn{9}{c}{$h=12$}\\
ols & 1.030 & 1.031 & 1.031 & 1.087 & 1.083 & 1.084 & 1.065 & 1.063 & 1.063\\
wls & \textbf{0.951} & \textbf{0.977} & \textbf{0.970} & \textbf{0.985} & \textbf{0.988} & \textbf{0.987} & \textbf{0.972} & \textbf{0.983} & \textbf{0.980}\\
shr & \em{0.970} & \em{1.015} & \em{1.003} & \em{1.004} & \em{1.026} & \em{1.020} & \em{0.991} & \em{1.022} & \em{1.014}\\
\addlinespace[0.3em]
\multicolumn{1}{c}{} & \multicolumn{9}{c}{$h=24$}\\
ols & 1.068 & 1.043 & 1.049 & 1.147 & 1.109 & 1.119 & 1.116 & 1.084 & 1.092\\
wls & \textbf{0.949} & \textbf{0.972} & \textbf{0.966} & \textbf{0.997} & \textbf{0.991} & \textbf{0.992} & \textbf{0.978} & \textbf{0.983} & \textbf{0.982}\\
shr & \em{0.984} & \em{0.998} & \em{0.994} & \em{1.033} & \em{1.022} & \em{1.025} & \em{1.014} & \em{1.013} & \em{1.013}\\
\bottomrule
\end{tabular}

	\endgroup
	\caption{Unemployment dataset. Average relative accuracy indexes ($gmRMSE$) where the reference forecasts are the base forecasts ($gmRMSE = 1$): lower values indicate better accuracy, bold highlights the best performance, italic indicates the second best. }
	\label{tab:gmRMSE_un}
\end{table}

At the shortest horizon ($h = 1$), all reconciliation methods improve upon the base forecasts, with the largest gains achieved by shr. For the upper-level aggregates, the $gmRMSE$ falls to 0.945 for \emph{Rates} and to 0.961 for \emph{All}, while wls obtains 0.967 and 0.979, respectively. The ols also delivers consistent, though slightly smaller, improvements across most series. The reduction in forecast errors at short horizons confirms that the non-linear reconciliation methods effectively correct incoherencies while preserving the information in the base forecasts.

For medium horizons ($h = 12$), the relative ranking of the methods changes. The wls approach becomes the most accurate overall, with $gmRMSE$ values around 0.97--0.98 for \emph{All} series, while shr is still competitive for the unemployment rate but has smaller or no improvements for the others variable (\emph{Others}). The ols reconciliation performs similarly to shr at this range, but less efficiently than wls. At the longest horizon ($h = 24$), the wls reconciliation consistently achieves the best results across all variable groups and aggregation levels. The $gmRMSE$ ranges between 0.949 and 0.982, while both shr and ols remain close to one but continue to outperform the base forecasts. These results confirm that, even in medium- and long-term settings, the proposed non-linear reconciliation methods maintain superior accuracy and internal coherence relative to independent forecasts.

\begin{figure}[tb]
	\centering
    \includegraphics[width = \linewidth]{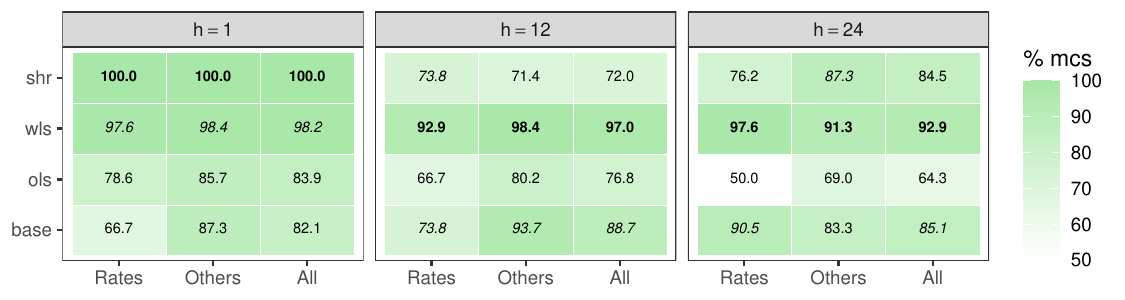}
    \caption{Unemployment dataset. Model Confidence Set results at 95\% threshold using squared loss: percentage of times each forecasting approach is included in the best confidence model set for different forecast horizon ($h$=1, 12, and 24) and for the rates (Rates), the labour force and the number of employed and unemployed individuals together (Others), and for all the variables (All). Values in {bold} indicate the method with the highest inclusion rate. Higher percentages indicate more frequent inclusion in the best model set, reflecting greater relative forecast accuracy.}
	\label{tab:MCS_MSE_un}
\end{figure}

\begin{figure}[!t]
	\centering
	\includegraphics[width = \linewidth]{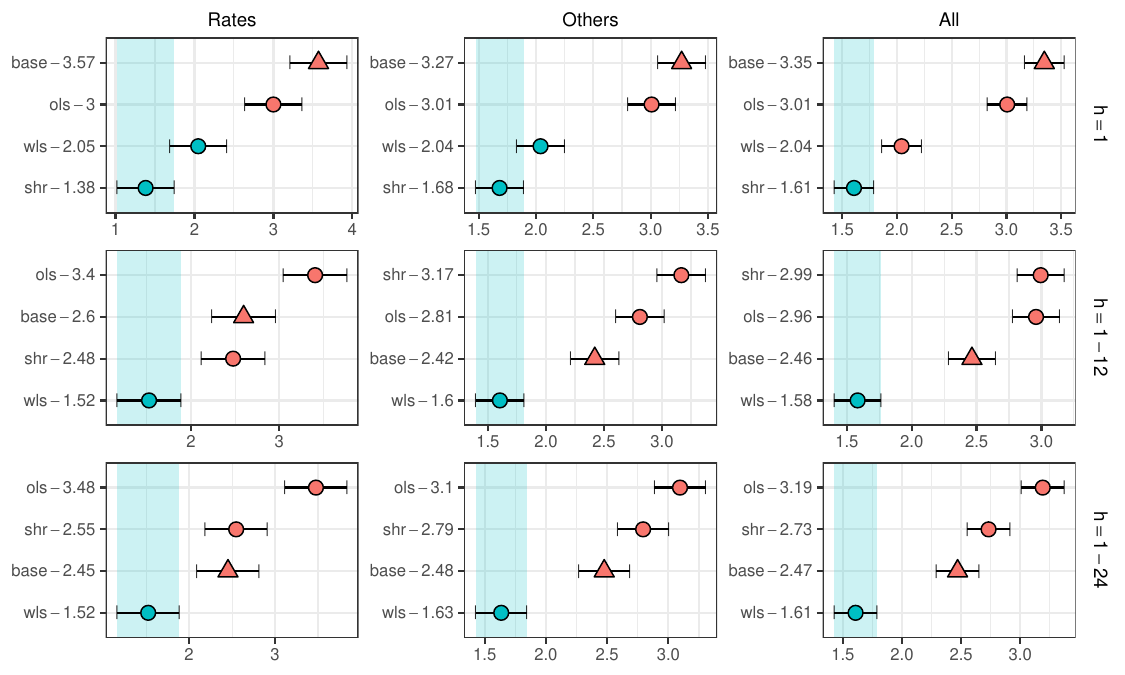}
	\caption{Unemployment dataset. MCB Nemenyi test for the Australian electricity generation dataset using the RMSE at different forecast horizon ($h$=1, 12, and 24) for the rates (Rates), the labour force and the number of employed and unemployed individuals together (Others), and for all the variables (All). In each panel, the Friedman test $p$-value is reported in the lower-right corner. The mean rank of each approach is shown to the right of its name. Statistically significant differences in performance are indicated if the intervals of two forecast reconciliation procedures do not overlap. Thus, approaches that do not overlap with the green interval are considered significantly worse than the best, and vice versa.}
	\label{fig:MCB_RMSE_unemp}
\end{figure}

The Diebold--Mariano tests (see supplementary material) indicate that the improvements obtained by shr and wls over the base forecasts are statistically significant at the 5\% level for most series, particularly for the national unemployment rate and the larger states. These findings are supported by the Model Confidence Set in \autoref{tab:MCS_MSE_un}: for $h=1$, shr is included in the best model set  almost always, with wls a close second; at $h=12$ and $h=24$ the wls reconciliation is most frequently selected in the best set, while shr is included less often. Rankings from the MCB Nemenyi procedure tell the same story: shr and wls are always better than the base and ols forecasts, as shown in \autoref{fig:MCB_RMSE_unemp}.

Overall, the evidence demonstrates that non-linearly constrained forecast reconciliation significantly improves both the accuracy and coherence of labour market forecasts, with consistent benefits across horizons and aggregation levels, most notably for the unemployment rate. In general, the shr approach is more effective for short-term forecasts, while wls approach for medium- and long-term planning. 

\section{Conclusion}\label{sec:conclusion}

Forecast reconciliation has been extended to the setting of non-linear constraints by considering projections to the coherent space, in a potentially (transformed) set of coordinates. This is achieved by framing reconciliation as an optimisation problem where the Lagrangian can be minimised with standard numerical methods. For linear constraints, a projection is guaranteed to improve forecast accuracy in a mean squared sense; the same result does not always hold for non-linear constraints. Accordingly, we develop theory in this setting by constructing a ball around the reconciled forecast such that for any true observation lying within the ball, forecast reconciliation improves forecast accuracy. The radius of this ball, and accordingly the probability that reconciliation does improve base forecasts depends on multiple factors. These include the curvature of constraints, the distance of base forecasts from the coherent space and the distance of reconciled forecasts from the mass of the true underlying data generating process. For the special case of convexity, additional results are derived.

The theoretical results are demonstrated in a simulated setting, and the new reconciliation methodologies are applied to two datasets in mortality and labour economics. The simulations verify the conclusions drawn from the theoretical analysis. It is also worth noting that even in the worst case simulated scenarios (high curvature, close to coherent base forecasts biased towards the epigraph of a convex constraint), reconciliation methods still improve on base forecasts with probabilities on the order of 80\%--90\%. Furthermore in the empirical studies, reconciliation methods significantly improve on base forecasts as well as on other competitive benchmarks. Improvements due to reconciliation are better when a weighting matrix is used in the optimisation, whether that weighting matrix is diagonal or set to an estimate of in-sample forecast error covariance.

There are a number of avenues for open research. In particular, only the distance reducing properties of projections are investigated here. A non-linear analogue to MinT is implemented and shown to have good performance in the empirical studies. Extending the results of \cite{wickramasuriya2019} to the the non-linear case would provide important theoretical underpinnings to our empirical work. Furthermore, in the linear setting, \cite{panagiotelis2021} consider mappings to the coherent subspace that are more general than projections. It would be worthwhile to investigate whether such approaches could be developed for non-linear constraints and whether they would outperform projection based approaches. Finally, theoretically deriving appropriate prediction intervals for forecasts projected onto surfaces is another challenge to be explored for non-linearly constrained forecast reconciliation.

\section*{Acknowledgments}
\phantomsection\addcontentsline{toc}{section}{Acknowledgments}

Professor Panagiotelis receives funding from the Australian Research Council Discovery Project scheme (project number DP250100702). Dr Li receives funding from the Australian Research Council Discovery Project scheme (project number DP220100090). Professor Di Fonzo and Dr Girolimetto receive funding from the project PRIN 2022 “PRICE: A New Paradigm for High-Frequency Finance" (project number 2022C799SX)

\appendix
\vspace*{-0.05cm}

\section*{Appendix}

\begingroup
\vspace*{-0.5cm}
\begin{algorithm}[H]
	\small
	\caption{SLSQP for non-linearly constrained forecast reconciliation}
	\label{alg:slsqp-nlcr}
	\begin{algorithmic}[1]
		\Require Base forecasts $\yhat \in \mathbb{R}^n$, constraints function $\gvet(\yvet) = \Zerovet$, covariance matrix $\widehat{\Wvet}\succ 0$, tolerances $\varepsilon_{1}$ and $\varepsilon_{2}$, step length $\alpha_k \in (0,1]$, max iterations $K$
		\State Choose initial point $\yvet_0$ (e.g., $\yvet_0 \gets \yhat$); set $k \gets 0$ and $\Bvet_0 \gets \Ivet_n$
		\While{$k < K$}
		\State Evaluate objective and gradient functions
		\[
		f(\yvet_k) \gets \tfrac{1}{2}(\yvet_k-\yhat)^\top \widehat{\Wvet}^{-1}(\yvet_k-\yhat)
		\quad \mathrm{and} \quad  \nabla f(\yvet_k) \gets \widehat{\Wvet}^{-1}(\yvet_k-\yhat)
		\]
		\State Evaluate constraints $\bm{g}(\yvet_k)$ and Jacobian $\nabla \bm{g}(\yvet_k)$
		\State Solve the quadratic subproblem to obtain search direction $\dvet_k$:
		\[
		\begin{aligned}
			\min_{\dvet \in \mathbb{R}^n}\quad & \tfrac{1}{2}\,\dvet^\top \Bvet_k\,\dvet \;+\; \nabla f(\yvet_k)^\top \dvet \\
			\text{s.t.}\quad & \bm{g}(\yvet_k) + \nabla \bm{g}(\yvet_k)\,\dvet = \mathbf{0}.
		\end{aligned}
		\]
		\State Update iterate: $\yvet_{k+1} \gets \yvet_k + \alpha_k \dvet_k$
		\State Compute $\Bvet_{k+1}$ from $\Bvet_k$ via a BFGS update
		\If{$\|\nabla f(\yvet_{k+1}) + \nabla \bm{g}(\yvet_{k+1})^\top \lambdavet_{k+1}\|\le \varepsilon_{1}$ and $\|\bm{g}(\yvet_{k+1})\|\le \varepsilon_{2}$}
		\State \textbf{break}
		\EndIf
		\State $k \gets k+1$
		\EndWhile
		\State \Return Reconciled forecasts $\ytilde \gets \yvet_{k}$
	\end{algorithmic}
\end{algorithm}
\endgroup

\phantomsection\addcontentsline{toc}{section}{References}

\bibliographystyle{CUP}
\bibliography{mybibfile.bib}

\end{document}


\def\spacingset#1{\renewcommand{\baselinestretch}{#1}\small\normalsize}
\spacingset{1.1}

\thispagestyle{empty}\clearpage\maketitle

\ifnum\blind=1
{
\begingroup
\let\thefootnote\relax\footnotetext{\raggedright Emails: \href{mailto:daniele.girolimetto@unipd.it}{\texttt{daniele.girolimetto@unipd.it}}, \href{mailto:anastasios.panagiotelis@monash.edu}{\texttt{anastasios.panagiotelis@monash.edu}}, \href{mailto:difonzo@stat.unipd.it}{\texttt{difonzo@stat.unipd.it}}, and \href{mailto:han.li@unimelb.edu.au}{\texttt{han.li@unimelb.edu.au}}}
\endgroup
}\fi

\appendix
\tableofcontents
\newpage
\renewcommand\thefigure{\thesection.\arabic{figure}}
\renewcommand\thetable{\thesection.\arabic{table}}

\section{Simulations -- Extended results}\label{sec:app_sim}

\begin{figure}[H]
    \renewcommand\thefigure{A.1}
	\centering
	\includegraphics[width=\textwidth]{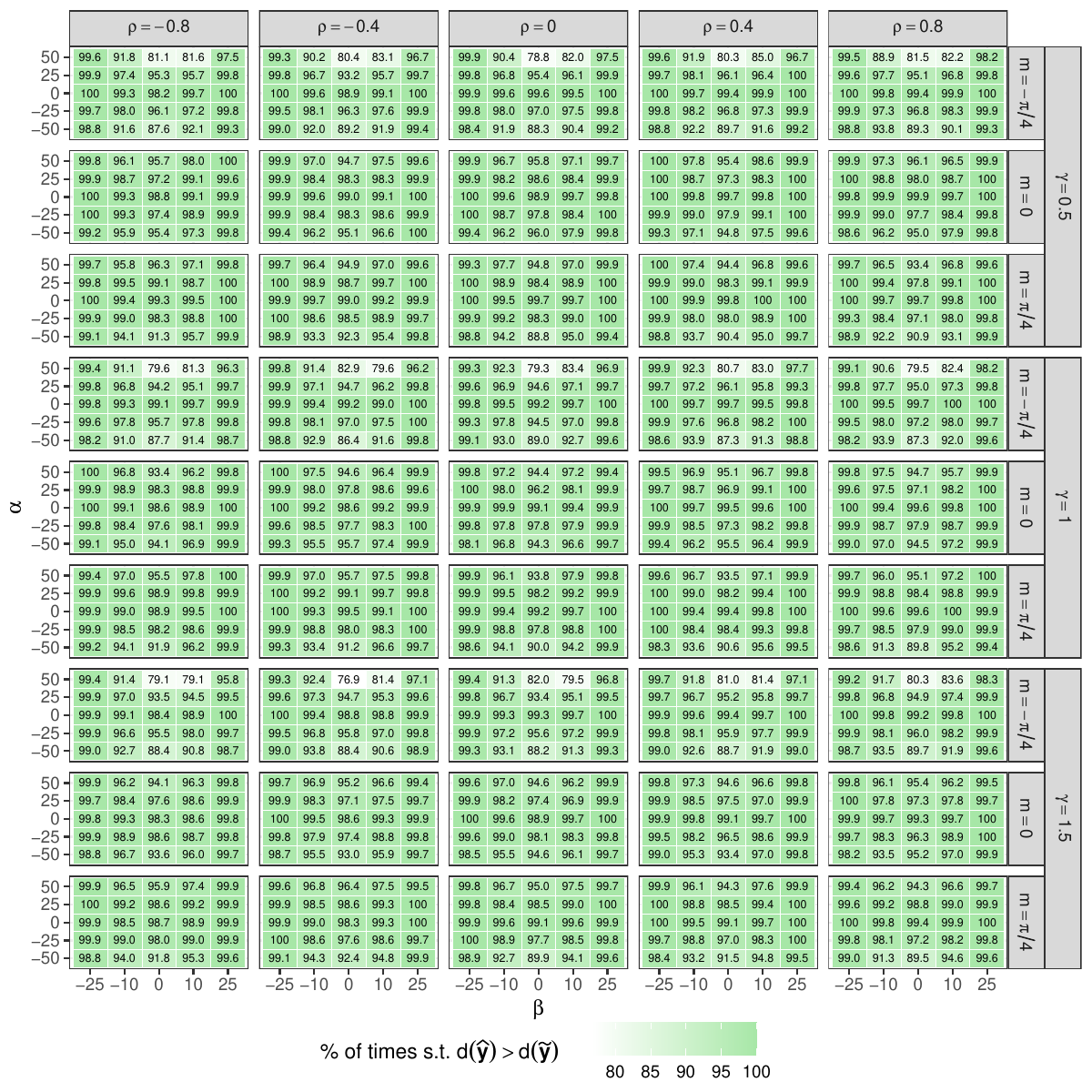}
	\caption{Simulation 2 results showing the proportion of times that reconciled forecasts are closer to the target value than base forecasts. Results are shown for the complete set of parameters.}
	\label{fig:res_sim2_all}
\end{figure}

\section{Mortality data -- Extended results}\label{sec:app1}

\begin{table}[H]
	\centering
	\begingroup
	\spacingset{1.1}
	\setlength{\tabcolsep}{4pt}
	\small
	
\begin{tabular}[t]{c|c|ccccccccc}
\toprule
\multicolumn{1}{c}{\textbf{}} & \multicolumn{1}{c}{\textbf{}} & \multicolumn{9}{c}{\textbf{Census Division}} \\
\textbf{Age} & \textbf{USA} & ENCD & ESCD & MAD & MD & NED & PD & SAD & WNCD & WSCD\\
\midrule
{}[0,5] & $\bm{<0.001}$ & $\bm{<0.001}$ & $\bm{<0.001}$ & $\bm{<0.001}$ & \textbf{0.046} & \textbf{0.004} & $\bm{<0.001}$ & $\bm{<0.001}$ & 0.051 & $\bm{<0.001}$\\
(5,10] & \textbf{0.009} & 0.441 & 0.413 & 0.818 & \textbf{0.011} & \textbf{0.001} & \textbf{0.017} & 0.578 & \textbf{0.004} & 0.323\\
(10,15] & \textbf{0.001} & 0.196 & \textbf{0.004} & 0.272 & 0.297 & 0.089 & \textbf{0.017} & 0.068 & 0.493 & \textbf{0.019}\\
(15,20] & \textbf{0.002} & \textbf{0.040} & 0.316 & 0.062 & 0.085 & \textbf{0.025} & \textbf{0.001} & 0.152 & 0.168 & 1.000\\
(20,25] & \textbf{0.010} & $\bm{<0.001}$ & 0.347 & \textbf{0.036} & \textbf{0.040} & 0.297 & \textbf{0.019} & \textbf{0.009} & 0.313 & 0.062\\
(25,30] & $\bm{<0.001}$ & $\bm{<0.001}$ & \textbf{0.001} & \textbf{0.006} & \textbf{0.002} & \textbf{0.010} & \textbf{0.002} & $\bm{<0.001}$ & \textbf{0.006} & \textbf{0.016}\\
(30,35] & $\bm{<0.001}$ & \textbf{0.001} & \textbf{0.002} & $\bm{<0.001}$ & \textbf{0.002} & $\bm{<0.001}$ & $\bm{<0.001}$ & $\bm{<0.001}$ & 0.067 & $\bm{<0.001}$\\
(35,40] & $\bm{<0.001}$ & \textbf{0.028} & $\bm{<0.001}$ & $\bm{<0.001}$ & 0.051 & \textbf{0.001} & $\bm{<0.001}$ & \textbf{0.001} & \textbf{0.022} & \textbf{0.013}\\
(40,45] & \textbf{0.004} & $\bm{<0.001}$ & $\bm{<0.001}$ & $\bm{<0.001}$ & \textbf{0.022} & \textbf{0.002} & 1.000 & $\bm{<0.001}$ & \textbf{0.002} & \textbf{0.002}\\
(45,50] & $\bm{<0.001}$ & $\bm{<0.001}$ & $\bm{<0.001}$ & 0.845 & $\bm{<0.001}$ & \textbf{0.012} & \textbf{0.010} & $\bm{<0.001}$ & $\bm{<0.001}$ & $\bm{<0.001}$\\
(50,55] & $\bm{<0.001}$ & $\bm{<0.001}$ & $\bm{<0.001}$ & $\bm{<0.001}$ & $\bm{<0.001}$ & \textbf{0.015} & $\bm{<0.001}$ & $\bm{<0.001}$ & $\bm{<0.001}$ & $\bm{<0.001}$\\
(55,60] & \textbf{0.001} & $\bm{<0.001}$ & $\bm{<0.001}$ & \textbf{0.007} & \textbf{0.001} & 0.199 & $\bm{<0.001}$ & \textbf{0.002} & \textbf{0.005} & \textbf{0.002}\\
(60,65] & \textbf{0.001} & \textbf{0.019} & $\bm{<0.001}$ & 0.350 & \textbf{0.008} & \textbf{0.001} & $\bm{<0.001}$ & $\bm{<0.001}$ & $\bm{<0.001}$ & $\bm{<0.001}$\\
(65,70] & \textbf{0.001} & $\bm{<0.001}$ & 1.000 & 0.105 & $\bm{<0.001}$ & $\bm{<0.001}$ & \textbf{0.005} & \textbf{0.013} & $\bm{<0.001}$ & \textbf{0.030}\\
(70,75] & $\bm{<0.001}$ & $\bm{<0.001}$ & 1.000 & \textbf{0.005} & $\bm{<0.001}$ & $\bm{<0.001}$ & $\bm{<0.001}$ & $\bm{<0.001}$ & 1.000 & \textbf{0.002}\\
(75,80] & $\bm{<0.001}$ & $\bm{<0.001}$ & $\bm{<0.001}$ & \textbf{0.007} & $\bm{<0.001}$ & $\bm{<0.001}$ & $\bm{<0.001}$ & $\bm{<0.001}$ & 1.000 & $\bm{<0.001}$\\
(80,85] & $\bm{<0.001}$ & \textbf{0.007} & $\bm{<0.001}$ & \textbf{0.030} & $\bm{<0.001}$ & \textbf{0.003} & \textbf{0.004} & $\bm{<0.001}$ & $\bm{<0.001}$ & $\bm{<0.001}$\\
\bottomrule
\end{tabular}

	\endgroup	
	\caption{Mortality data. Diebold-Mariano test results (base vs. $shr$) using squared loss. The test is conducted for each age group and across all census divisions as well as the USA total. Lower p-values indicate stronger evidence that $shr$ is more accurate than the base forecast. In the table, values in \textbf{bold} indicate statistically significant superiority of $shr$ at the 5\% level ($p<0.05$), italic values indicate second-best significance, and values in red correspond to $p$-values greater than 0.1, suggesting no significant difference from the base forecast.}
	\label{tab:DM_MSE_mortality}
\end{table}

\begin{table}[H]
	\centering
	\begingroup
	\spacingset{1.1}
	\small
	
\begin{tabular}[t]{l|ccc|ccc|ccc}
\toprule
\multicolumn{1}{c}{\textbf{}} & \multicolumn{3}{c}{\textbf{Rates}} & \multicolumn{3}{c}{\textbf{Others}} & \multicolumn{3}{c}{\textbf{All}} \\
\textbf{App.} & $\delta =95$ & $\delta =90$ & $\delta =75$ & $\delta =95$ & $\delta =90$ & $\delta =75$ & $\delta =95$ & $\delta =90$ & $\delta =75$\\
\midrule
\addlinespace[0.3em]
\multicolumn{1}{c}{} & \multicolumn{9}{c}{$h=1$}\\
base & 22.4 & 15.3 & 10.6 & 93.5 & \textbf{88.2} & \em{\textbf{75.0}} & 69.8 & 63.9 & 53.5\\
bu & 50.6 & 41.2 & 30.0 & 92.4 & 87.1 & \em{\textbf{75.0}} & 78.4 & 71.8 & 60.0\\
LH & 22.9 & 14.7 & 10.6 &  &  &  & 22.9 & 14.7 & 10.6\\
ols & 94.7 & 88.2 & 70.0 & 95.0 & 85.3 & 66.5 & 94.9 & 86.3 & 67.6\\
wls & \em{98.2} & \em{91.8} & \em{77.1} & \em{95.6} & 86.5 & 69.7 & \em{96.5} & \em{88.2} & \em{72.2}\\
shr & \textbf{100.0} & \textbf{99.4} & \textbf{94.7} & \textbf{96.5} & \em{87.9} & 74.1 & \textbf{97.6} & \textbf{91.8} & \textbf{81.0}\\
\addlinespace[0.3em]
\multicolumn{1}{c}{} & \multicolumn{9}{c}{$h=5$}\\
base & 77.6 & 71.2 & 55.3 & 92.6 & 88.2 & 76.5 & 87.6 & 82.5 & 69.4\\
bu & 77.6 & 71.8 & 54.7 & 92.1 & 87.6 & 75.6 & 87.3 & 82.4 & 68.6\\
LH & 77.1 & 73.5 & 60.0 &  &  &  & 77.1 & 73.5 & 60.0\\
ols & 92.9 & \em{89.4} & 73.5 & 96.5 & 91.5 & 76.8 & 95.3 & 90.8 & 75.7\\
wls & \em{94.1} & \em{89.4} & \em{77.6} & \em{97.9} & \em{93.2} & \em{80.6} & \em{96.7} & \em{92.0} & \em{79.6}\\
shr & \textbf{95.9} & \textbf{91.8} & \textbf{79.4} & \textbf{98.2} & \textbf{94.4} & \textbf{85.3} & \textbf{97.5} & \textbf{93.5} & \textbf{83.3}\\
\addlinespace[0.3em]
\multicolumn{1}{c}{} & \multicolumn{9}{c}{$h=10$}\\
base & 81.8 & \em{76.5} & \em{65.9} & 86.2 & \em{80.6} & \textbf{69.4} & 84.7 & \textbf{79.2} & \em{68.2}\\
bu & 70.6 & 64.7 & 50.0 & 85.6 & 79.4 & 67.6 & 80.6 & 74.5 & 61.8\\
LH & 80.6 & \textbf{78.2} & \textbf{69.4} &  &  &  & 80.6 & 78.2 & \textbf{69.4}\\
ols & 81.8 & 73.5 & 59.4 & 87.4 & \textbf{81.5} & \em{68.2} & 85.5 & \em{78.8} & 65.3\\
wls & \em{82.4} & 73.5 & 62.4 & \textbf{88.8} & \em{80.6} & 67.1 & \textbf{86.7} & 78.2 & 65.5\\
shr & \textbf{82.9} & 74.7 & 60.0 & \em{88.2} & 80.0 & 67.6 & \em{86.5} & 78.2 & 65.1\\
\bottomrule
\end{tabular}

	\endgroup
	\caption{Mortality data. Model Confidence Set results with different
		thresholds ($\delta \in \{95\%, 90\%, 75\%\}$) using squared loss: percentage of times each forecasting approach is included in the best confidence model set for different forecast horizons ($h$=1, 5, and 10) and for the rates (Rates), the population exposures and death counts together (Others), and for all the variables (All). Values in \textbf{bold} indicate the method with the highest inclusion rate for a given metric and horizon. Higher percentages indicate more frequent inclusion in the best model set, reflecting greater relative forecast accuracy.}
	\label{tab:MCS_MSE_mortality_app}
\end{table}

\section{Unemployment data -- Extended results}\label{sec:app2}

\begin{table}[H]
\renewcommand\thetable{C.1}
	\centering
	\begingroup
	\spacingset{1.1}
	\footnotesize
	
\begin{tabular}[t]{l|ccc}
\toprule
\multicolumn{1}{c}{\em{}} & \multicolumn{3}{c}{\em{Square err.}} \\
Variable & ols & wls & shr\\
\midrule
Australia & \textbf{0.009} & 0.155 & \textbf{0.001}\\
\addlinespace
Females & 0.416 & 0.261 & 0.078\\
Males & \textbf{0.007} & \textbf{0.046} & \textbf{0.010}\\
\addlinespace
ACT & 0.999 & \textbf{0.016} & \textbf{0.004}\\
NSW & 0.385 & 0.090 & 0.130\\
NT & 0.997 & \textbf{0.035} & \textbf{0.021}\\
QLD & \textbf{0.002} & \textbf{0.003} & \textbf{0.009}\\
SA & 0.526 & 0.338 & 0.114\\
TAS & 0.998 & \textbf{0.037} & 0.190\\
VIC & 0.054 & \textbf{0.004} & \textbf{0.043}\\
WA & \textbf{0.007} & $\bm{<0.001}$ & $\bm{<0.001}$\\
\addlinespace
15-24 & \textbf{0.027} & \textbf{0.010} & \textbf{0.004}\\
25-34 & \textbf{0.020} & \textbf{0.002} & \textbf{0.012}\\
35-44 & \textbf{0.018} & \textbf{0.005} & \textbf{0.003}\\
45-54 & \textbf{0.001} & $\bm{<0.001}$ & $\bm{<0.001}$\\
55-64 & \textbf{0.041} & \textbf{0.002} & $\bm{<0.001}$\\
\addlinespace
NSW Females & 0.182 & \textbf{0.008} & 0.094\\
VIC Females & 0.060 & \textbf{0.001} & \textbf{0.016}\\
QLD Females & 0.271 & \textbf{0.021} & \textbf{0.018}\\
SA Females & 0.101 & \textbf{0.002} & \textbf{0.013}\\
WA Females & 0.438 & 0.150 & 0.168\\
TAS Females & 0.977 & \textbf{0.008} & 0.067\\
NT Females & 0.998 & 0.613 & 0.077\\
ACT Females & 1.000 & 0.128 & \textbf{0.026}\\
NSW Males & \textbf{0.040} & \textbf{0.003} & \textbf{0.041}\\
VIC Males & 0.248 & 0.107 & 0.179\\
QLD Males & $\bm{<0.001}$ & $\bm{<0.001}$ & \textbf{0.002}\\
SA Males & 0.673 & \textbf{0.037} & 0.234\\
WA Males & \textbf{0.012} & \textbf{0.005} & \textbf{0.007}\\
TAS Males & 0.999 & 0.148 & 0.608\\
NT Males & 1.000 & 0.639 & 0.704\\
ACT Males & 1.000 & 0.625 & 0.720\\
\addlinespace
15-24 Females & 0.209 & \textbf{0.028} & \textbf{0.049}\\
25-34 Females & 0.095 & \textbf{0.006} & 0.142\\
35-44 Females & 0.112 & \textbf{0.005} & 0.066\\
45-54 Females & \textbf{0.008} & $\bm{<0.001}$ & $\bm{<0.001}$\\
55-64 Females & 0.699 & \textbf{0.042} & 0.051\\
15-24 Males & 0.139 & \textbf{0.009} & \textbf{0.005}\\
25-34 Males & \textbf{0.049} & \textbf{0.008} & \textbf{0.024}\\
35-44 Males & 0.301 & 0.110 & 0.184\\
45-54 Males & 0.135 & \textbf{0.003} & \textbf{0.019}\\
55-64 Males & 0.208 & \textbf{0.001} & \textbf{0.003}\\
\bottomrule
\end{tabular}

	\endgroup
	\caption{Unemployment data. Diebold-Mariano test results (base vs. $shr$) using squared loss. The test is conducted for each variable at forecast horizon $h = 1$. Lower p-values indicate stronger evidence that $shr$ is more accurate than the base forecast. In the table, values in \textbf{bold} indicate statistically significant superiority of $shr$ at the 5\% level ($p<0.05$), italic values indicate second-best significance, and values in red correspond to $p$-values greater than 0.1, suggesting no significant difference from the base forecast.}
	\label{tab:DM_MSE_un}
\end{table}

\begin{figure}[H]
\renewcommand\thefigure{C.1}
	\centering
	\includegraphics[width = \linewidth]{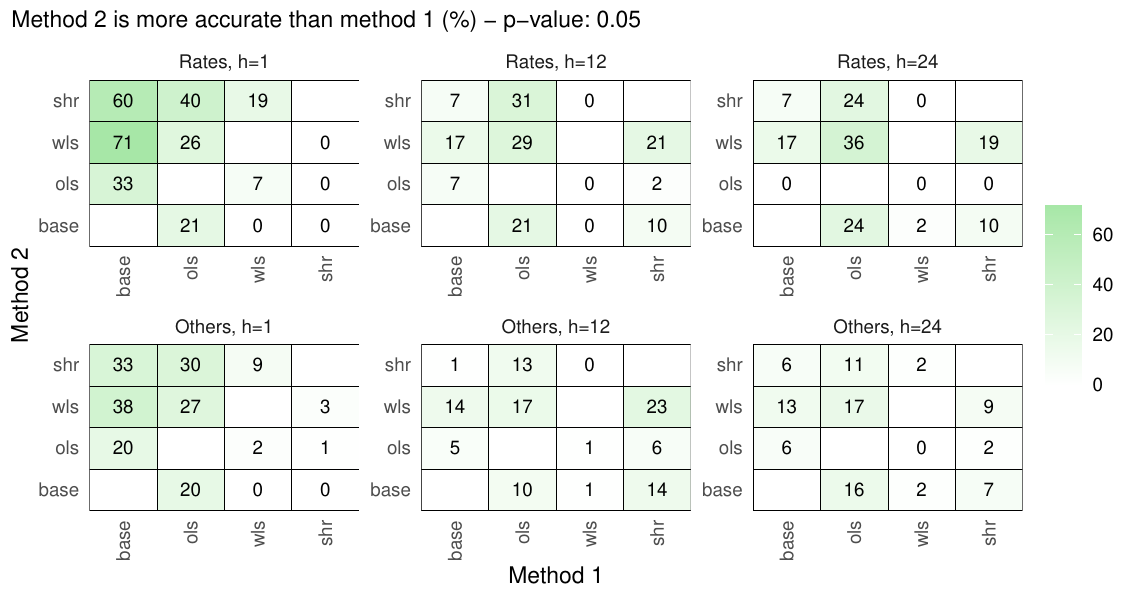}
	\caption{Unemployment data. Pairwise DM-test results evaluated using squared loss across different forecast horizons ($h$=1, 12, and 24) and for the rates (Rates) and the labour force and the number of employed and unemployed individuals together (Others). Each cell reports the number of times (in \%) the forecasting approach in the row statistically outperforms ($p-values<0.05$) the approach in the column.}
	\label{fig:DM_RMSE_unemp}
\end{figure}

\begin{table}[H]
\renewcommand\thetable{C.2}
	\centering
	\begingroup
	\spacingset{1.1}
	\small
	
\begin{tabular}[t]{l|ccc|ccc|ccc}
\toprule
\multicolumn{1}{c}{\textbf{}} & \multicolumn{3}{c}{\textbf{Rates}} & \multicolumn{3}{c}{\textbf{Others}} & \multicolumn{3}{c}{\textbf{All}} \\
\textbf{App.} & $\delta = 95$ & $\delta = 90$ & $\delta = 75$ & $\delta = 95$ & $\delta = 90$ & $\delta = 75$ & $\delta = 95$ & $\delta = 90$ & $\delta = 75$\\
\midrule
\addlinespace[0.3em]
\multicolumn{1}{c}{} & \multicolumn{9}{c}{$h=1$}\\
base & 66.7 & 42.9 & 26.2 & 87.3 & 74.6 & 53.2 & 82.1 & 66.7 & 46.4\\
ols & 78.6 & 73.8 & 61.9 & 85.7 & 79.4 & 69.0 & 83.9 & 78.0 & 67.3\\
wls & \em{97.6} & \em{90.5} & \em{73.8} & \em{98.4} & \em{94.4} & \em{81.7} & \em{98.2} & \em{93.5} & \em{79.8}\\
shr & \textbf{100.0} & \textbf{100.0} & \textbf{100.0} & \textbf{100.0} & \textbf{100.0} & \textbf{99.2} & \textbf{100.0} & \textbf{100.0} & \textbf{99.4}\\
\addlinespace[0.3em]
\multicolumn{1}{c}{} & \multicolumn{9}{c}{$h=12$}\\
base & \em{73.8} & \em{71.4} & \em{64.3} & \em{93.7} & \em{84.1} & \em{66.7} & \em{88.7} & \em{81.0} & \em{66.1}\\
ols & 66.7 & 64.3 & 40.5 & 80.2 & 76.2 & 57.1 & 76.8 & 73.2 & 53.0\\
wls & \textbf{92.9} & \textbf{90.5} & \textbf{83.3} & \textbf{98.4} & \textbf{92.9} & \textbf{84.9} & \textbf{97.0} & \textbf{92.3} & \textbf{84.5}\\
shr & \em{73.8} & 69.0 & 59.5 & 71.4 & 63.5 & 42.1 & 72.0 & 64.9 & 46.4\\
\addlinespace[0.3em]
\multicolumn{1}{c}{} & \multicolumn{9}{c}{$h=24$}\\
base & \em{90.5} & \em{83.3} & \em{69.0} & 83.3 & \em{78.6} & 61.9 & \em{85.1} & \em{79.8} & 63.7\\
ols & 50.0 & 31.0 & 21.4 & 69.0 & 56.3 & 34.9 & 64.3 & 50.0 & 31.5\\
wls & \textbf{97.6} & \textbf{90.5} & \textbf{78.6} & \textbf{91.3} & \textbf{83.3} & \textbf{69.0} & \textbf{92.9} & \textbf{85.1} & \textbf{71.4}\\
shr & 76.2 & 69.0 & 54.8 & \em{87.3} & \em{78.6} & \em{67.5} & 84.5 & 76.2 & \em{64.3}\\
\bottomrule
\end{tabular}

	\endgroup
	\caption{Unemployment data. Model Confidence Set results with different
		thresholds ($\delta \in \{95\%, 90\%, 75\%\}$) using squared loss: percentage of times each forecasting approach is included in the best confidence model set for different forecast horizon ($h$=1, 12, and 24) and for the rates (Rates), the labour force and the number of employed and unemployed individuals together (Others), and for all the variables (All). Values in \textbf{bold} indicate the method with the highest inclusion rate. Higher percentages indicate more frequent inclusion in the best model set, reflecting greater relative forecast accuracy.}
	\label{tab:MCS_MSE_un_app}
\end{table}